\documentclass[fleqn,usenatbib]{mnras}

\usepackage{newtxtext,newtxmath}

\usepackage[T1]{fontenc}
\usepackage{ae,aecompl}

\usepackage{graphicx}	
\usepackage{amsmath}	
\usepackage{amssymb}	
\usepackage{multicol}        
\usepackage{bm}		
\usepackage{pdflscape}	
\usepackage{setspace} 
\usepackage{subfigure}
\usepackage{float}
\usepackage[british]{babel}
\usepackage{ulem} 
\usepackage{xcolor} 
\usepackage{multirow}

\title[Outliers in globular cluster 47\,Tuc]{\textit{Outliers:} multicolour photometry guiding the search for evolved binary systems in the globular cluster 47\,Tucanae}

\author[F. Campos]{Fab\'iola Campos$^{1,2}$\thanks{Contact e-mail: \href{mailto:fcampos@astro.as.utexas.edu}{fcampos@astro.as.utexas.edu}}\thanks{Based on observations made with the NASA/ESA Hubble Space Telescope, and obtained from the Hubble Legacy Archive, which is a collaboration between the Space Telescope Science Institute (STScI/NASA), the Space Telescope European Coordinating Facility (ST-ECF/ESA) and the Canadian Astronomy Data Centre (CADC/NRC/CSA).},
I. Pelisoli$^{2,3}$, S. Kamann$^{4,5}$,T.-O. Husser$^{4}$, S. Dreizler$^{4}$, A. Bellini$^{6}$,
\newauthor E. L. Robinson$^{1}$, D. Nardiello$^{7,8}$, G. Piotto$^{7,8}$, S. O. Kepler$^{2}$, A. G. Istrate$^{9}$, D. E. Winget$^{1}$,
\newauthor M. H. Montgomery$^{1}$, A. Dotter$^{10}$  
\\ 
$^{1}$ Department of Astronomy, University of Texas at Austin, 2515 Speedway, C1400, Austin, TX, 78712, USA\\ 
$^{2}$ Instituto de F\'{i}sica, Universidade Federal do Rio Grande do Sul, Av. Bento Gon\c{c}alves 9500, 91501-900, Porto Alegre, RS, Brazil\\
$^{3}$Institute for Physics and Astronomy, University of Potsdam, Karl Liebknecht-Str. 24/25, D-14476 Potsdam, Germany\\
$^{4}$Institute for Astrophysics, Georg-August-University, Friedrich-Hund-Platz 1, 37077 Gottingen, Germany\\
$^{5}$Astrophysics Research Institute, Liverpool John Moores University, 146 Brownlow Hill, Liverpool L3 5RF, United Kingdom\\
$^{6}$Space Telescope Science Institute, 3700 San Martin Dr., Baltimore, MD 21218, USA\\
$^{7}$Dipartimento di Fisica e Astronomia ``Galileo Galile'', Universit\`{a} di Padova, Vicolo dell'Osservatorio 3, I-35122, Padova, Italy\\
$^{8}$Istituto Nazionale di Astrofisica - Osservatorio Astronomico di Padova, Vicolo dell'Osservatorio 5, I-35122, Padova, Italy\\
$^{9}$Center for Gravitation, Cosmology, and Astrophysics, Dept. of Physics, Uni. of Wisconsin-Milwaukee, Milwaukee, WI, 53201, USA\\
$^{10}$Harvard-Smithsonian Center for Astrophysics, Cambridge, MA 02138, USA}

\date{Accepted XXX. Received YYY; in original form ZZZ}

\pubyear{2015}

\begin{document}
\label{firstpage}
\pagerange{\pageref{firstpage}--\pageref{lastpage}}
\maketitle

\begin{abstract}
We use Hubble Space Telescope multicolour photometry of the globular cluster 47\,Tucanae to uncover a population of 24 objects with no previous classification that are outliers from the single-star model tracks in the colour-magnitude diagram and yet are likely cluster members. By comparing those sources with evolutionary models and X-ray source catalogues, we were able to show that the majority of those sources are likely binary systems that do not have any X-ray source detected nearby, most possibly formed by a white dwarf and a main-sequence star and a small number of possible double-degenerate systems. 
\end{abstract}




\section{Introduction}

The cores of globular clusters (GCs) can reach extremely high stellar densities of $10^5M_{\odot}$pc$^{-3}$, making gravitational interactions and collisions among single stars and/or binaries to be fairly frequent in these environments \citep{Hut1992,Davies2002}. Such dynamical processes might result in the formation of binary systems through dissipative processes, such as tidal capture \citep{Fabian1975,McMillan1987}, and dissipationless stellar-dynamical processes \citep[e.g][]{Tanikawa2012}. However, dynamical processes might also lead to the destruction of wide binaries, hardening of close binaries and exchange interactions \citep{Heggie1975,Hut1983}. These exchange interactions, which are favoured in clusters with dense and/or large cores \citep{Hut1983a,Ivanova2006,Belloni2016,Hong2017}, are able to replace low mass companions by a more massive participant in an encounter, usually a white dwarf \citep{Shull1979}. While an encounter between a single star and a hard binary can result in the ejection of one star from the system or a collision, reducing the three-body system
into a binary \citep{McMillan1986}. Such processes deplete the cores of GCs of binaries, with the remaining systems showing period distributions very different from a Galactic field population \citep{Ivanova2005}.

The importance of the study of binary systems is widely acknowledged. The detection of gravitational radiation from binaries in GCs with very short periods can provide insights about the nature and evolution of GC \citep{Benacquista2013}. Binary systems provide the source of energy that supports GCs against gravothermal collapse \citep[e.g.][]{Goodman1989,Giersz2000}, strongly influencing the dynamical evolution of GCs \citep{Ivanova2005}. Monte Carlo simulations \citep{Heggie2006} indicate that  a cluster with no binaries goes into core collapse very quickly. However, the existence of even a small binary star population can deeply affect the core collapse \citep{Heggie1992}. Numerical simulations suggest that the present-day binary fractions in GCs are consistent with an initial binary fraction near to one \citep{Leigh2015}. And since the binary fraction in most GCs is significantly smaller than that in the field \citep{Milone2012}, and it seems to decrease as the cluster ages \citep[e.g.][]{Sollima2007}, the dynamical disruption of primordial binaries must have been efficient. 

The most well characterized population of binary stars in GCs is the main-sequence--main-sequence (MS--MS) binary systems. These systems populate the region in the colour-magnitude diagram (CMD) from the main-sequence until 0.753 magnitudes brighter than the overall main sequence \citep{Hurley1998}. However, the total fraction of binaries in GCs is much smaller than the one determined in the Galactic field \citep{Milone2012}. 

Binary systems that contain at least one star that is no longer on the main-sequence, i.e., evolved binary systems, can be roughly classified in two categories: \begin{enumerate}
\item the distance between the stars is large enough that the binary components evolve as quasi-single stars. Those systems can evolve, at first, to a binary composed by a main-sequence star plus a white dwarf (WD), neutron star (NS) or a black hole (BH), depending on the mass of the primary star. Later, if the systems are not disrupted, the secondary star evolves, and a double-degenerate system can be formed.
\item the binary components have little or no interaction during their main-sequence lifetimes. However, the stars can be sufficiently close to start binary interaction in the giant phases after the main-sequence, taking their evolution on a path different than the ones of single stars. Those binary systems are called stellar exotica \citep{Davies2002}. Among the stellar exotica are the cataclysmic variables \cite[CVs, e.g.][]{Knigge2011};  millisecond pulsars \citep[MSPs, e.g.][]{Lorimer2008}; low-mass X-ray binaries \citep[LMXBs, e.g.][]{Liu2007}; and extremely low mass WDs \citep[ELMs, e.g.][]{Brown2010}.
 \end{enumerate}

While some evolved binary systems in GCs can only be detected through thorough radial velocity variation analysis \citep[e.g., main-sequence--neutron-star, ][]{Giesers2018}, many of them can be uncovered due to their peculiar behaviour when compared to single stars. For example, stellar exotica are the main source of X--ray emissions in GCs \citep[CVs, LMXBs, UCXBs and MSPs,][]{Pooley2010, Nelemans2010, Henleywillis2018} and can be identified in X--ray surveys. The close--binary population containing a WD that is not currently interacting or that is transferring mass in extremely low rates, on the other hand, can only be detected with the use of multicolour photometry, mainly with the use of ultraviolet (UV) and visual CMDs. Confirmation on whether those sources are binaries can only be obtained with spectroscopy. 

In a search for new CVs in the core of the GC NGC\,6397, \cite{Cool1998} noted that in the CMD seven UV-bright sources are bluer than the main-sequence in the (F336W-F439W) colour. Four of these sources presented variability and are very close to or in the main-sequence in (F555W-F814W), placing them as CV candidates. The three remaining sources are still in the blue side of the main-sequence in (F555W-F814W); \cite{Cool1998} proposed they are He-core WDs resulting from binary evolution (currently known as ELMs), and they were the first publication to acknowledge the possible presence of those objects in GCs. Additional ELM candidates were identified in NGC\,6397 by \citet{Taylor2001} and \citet{Strickler2009} showed that the two dozen best  ELM candidates are strongly concentrated toward the cluster centre, implying the presence of unseen binary companions with an estimated average mass of about 1\,M$_{\odot}$, and argued that the companions are likely to be heavy white dwarfs.

More recently, \citet{Haurberg2010} identified populations that result from non-single stellar-evolution in the central region of the GC M\,15, including possible CVs and ELMs, by using photometric data from UV and blue filters. Among their data, they found a group of sources that appear in the gap between the main-sequence and the white dwarf cooling sequence in the UV CMD but lie on the main-sequence in the blue CMD. Unfortunately, even though X-ray measurements for M\,15 are available, \citet{Haurberg2010} could not use X-ray measurements to constrain the nature of those objects because the central region of M\,15 is dominated by two extremely bright X-ray sources (LMXBs), making X-ray information on dim objects unavailable. 

We began our search for those outliers with the globular cluster 47\,Tucanae (47\,Tuc), a metal rich and massive GC that is rather close to the Sun \citep[$\sim$4.3\,kpc, ][]{Heyl2017}. The high number of X--ray sources identified in 47\,Tuc \citep{Heinke2005}, along with near--ultraviolet and optical images from the Hubble Space Telescope (HST), allowed \citet{Sandoval2018} to identify 43 CVs and CV candidates in 47\,Tuc, the largest number ever found in a GC. They also found that the CVs in 47\,Tuc are more concentrated toward the cluster centre than the main-sequence-turnoff stars. However, no information on WD--MS and WD--WD binaries with no X-Ray emission detected has been reported. 

Using HST data obtained with the Wide Field Camera 3 (WFC3) and the Advanced Camera for Surveys (ACS), along with high-precision internal proper motions determined by \citet{Bellini2014}, we have identified a considerable number of non-previously catalogued outliers in the central part of 47\,Tuc, most of them with no X--ray source detected nearby. We show that most of those outliers are likely WD--MS binaries, while a few of them might be double-degenerate binaries, ELM candidates and CV candidates that were not previously reported. 

This paper is organised as follows: in Sect. \ref{phot} we give a brief description of all the data sources used in this paper and describe the analysis we performed with the photometric data. Discussion is presented in Sect. \ref{discussion}. Concluding remarks are given in Sect. \ref{conclusions}.

\section{Photometric data and analysis}
\label{phot}

\begin{figure}
\subfigure[\label{subfig1a}]{\includegraphics[width=0.3\textheight, trim={1cm 1.5cm 1cm 1.8cm},clip]{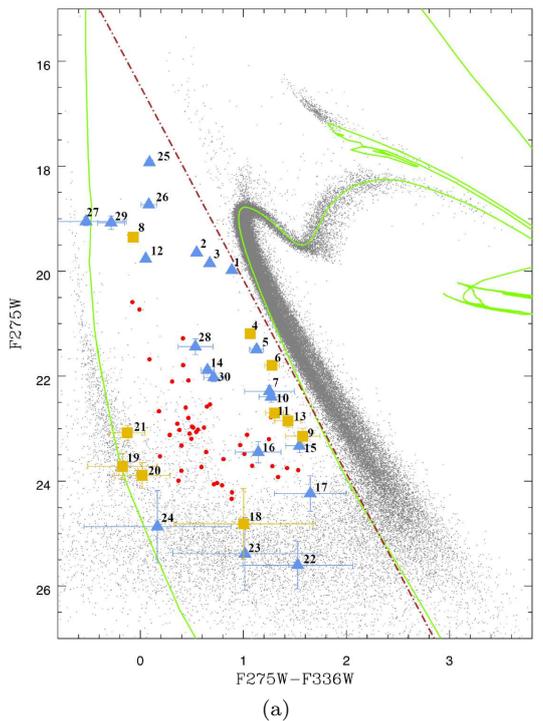}}
\subfigure[\label{subfig1b}]{\includegraphics[width=0.3\textheight, trim={1cm 2.5cm 1.3cm 3cm},clip]{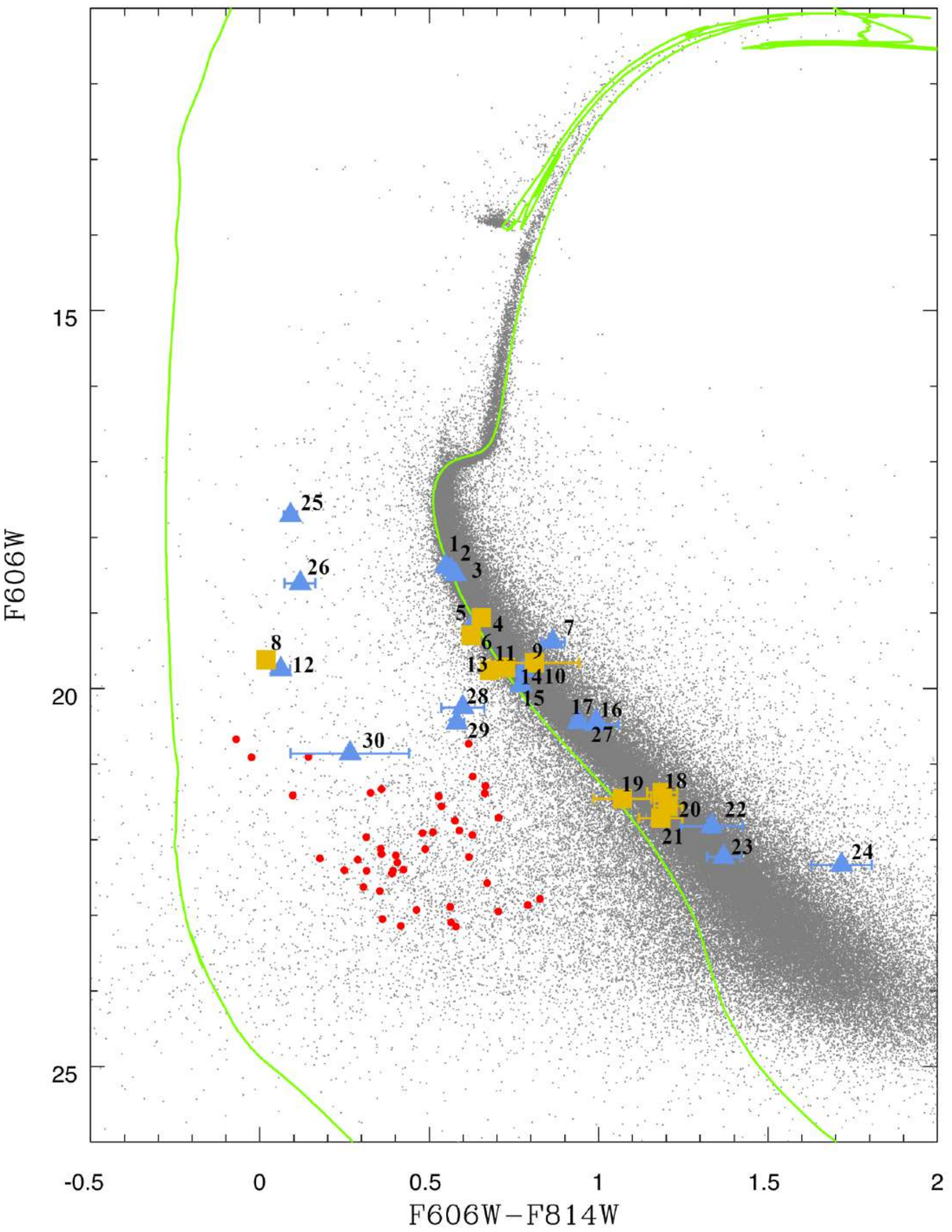}}
\caption{The CMDs (light grey) in F275W\,vs.\,F275W-F336W (panel a) and F606W\,vs.\,F606W-F814W (panel b). All sources that are in the bluer than the brown dot-dashed line in F275W\,vs.\,F275W-F336W, bluer than the main-sequence, and were cross-identified by \protect\citep{Bellini2014} are highlighted as follows: the light-blue squares denote most likely cluster members, the yellow triangles are possible cluster members, and the red circles are most likely SMC members. The green lines are isochrones with 11.5\,Gyr and \textit{Z}=0.003 calculated using MIST \protect\citep{Dotter2016,Choi2016}, from the low main-sequence down to white dwarfs with ${T}_\text{eff}\sim$23\,000\,K, and BaSTI \protect\citep{Salaris2010} for white dwarfs cooler than ${T}_\text{eff}\sim$23\,000\,K (Sect. \ref{cmdsection}). All the outliers with proper motion consistent with the one from the cluster have the uncertainties in their photometry shown in the CMD.}
 \label{fig1}
\end{figure}

We use the photometric catalogue from the final release of the treasury program ``The HST legacy survey of Galactic globular clusters: shedding UV light on their populations and formation'' (GO 13297) as described in \citet{Piotto2015,Nardiello2018}. With 131 orbits allocated in cycle 21 from HST, this treasury program obtained UV/blue WFC3/UVIS images with filters F275W, F336W and F438W for most of the clusters present in the existing F606W and F814W ACS/WFC database entitled ``An ACS Survey of Galactic Globular Cluster'' (PI Sarajedini, GO-10775, 134 orbits). 

\begin{hyphenrules}{british}
\hyphenation{colours}
\end{hyphenrules}

For 47\,Tuc, the catalogue of the final release from \citet{Piotto2015} and \citet{Nardiello2018} included multiple HST datasets (GO-9443 PI:King, GO-9281 PI:Grindlay, GO-10775 PI:Sarajedini, GO-11729 PI: Holtzman, GO-12971 PI: Richer, GO-12311 PI:Piotto) obtained between 2002 and 2012. That provides several images for the same filter in multiple epochs: 14 frames for F275W, 9 frames for F336W, 4 frames for F606W and 4 frames for F814W. \citet{Nardiello2018} analysed all the individual images simultaneously, so if a star was found in one filter in a given position, the routine measured the flux in all the other filters in the same position. The averaged flux in each filter was calculated and the RMS was determined and set as the scatter in the magnitude of a particular star.  

\citet{Nardiello2018} obtained photometry on filters F275W, F336W, F435W, F606W and F814W. This provides the two different colours necessary to search for stellar exotica using the same method applied by \citet{Cool1998} ( i.e., to look for the objects that are bluer than main-sequence in the UV/blue colours) and locate those objects on the visual CMD. 

In addition to \citet{Nardiello2018}'s data, we also used the photometric data available on the Hubble Legacy Archive (HLA) for the filters F658N (H$\alpha$) and F625W. This data was obtained in three different epochs as part of the program GO 9281.

In the upper panel of Fig. \ref{fig1} we show the F275W\,vs.\,F275W-F336W CMD (from now on referred as ``UV CMD''), we selected the sources bluer than the main-sequence in the UV CMD (bluer than the dash-dotted line in Fig \ref{fig1}). However, the region of the gap between the main-sequence and the white dwarf cooling sequence on the colour-magnitude diagram can be contaminated by foreground/background objects. Therefore, information on the cluster membership of the objects located in that area of the colour-magnitude diagram is imperative. 

\begin{figure}
\center
 \includegraphics[width=0.50\textwidth, trim={0 0 37cm 37cm},clip]{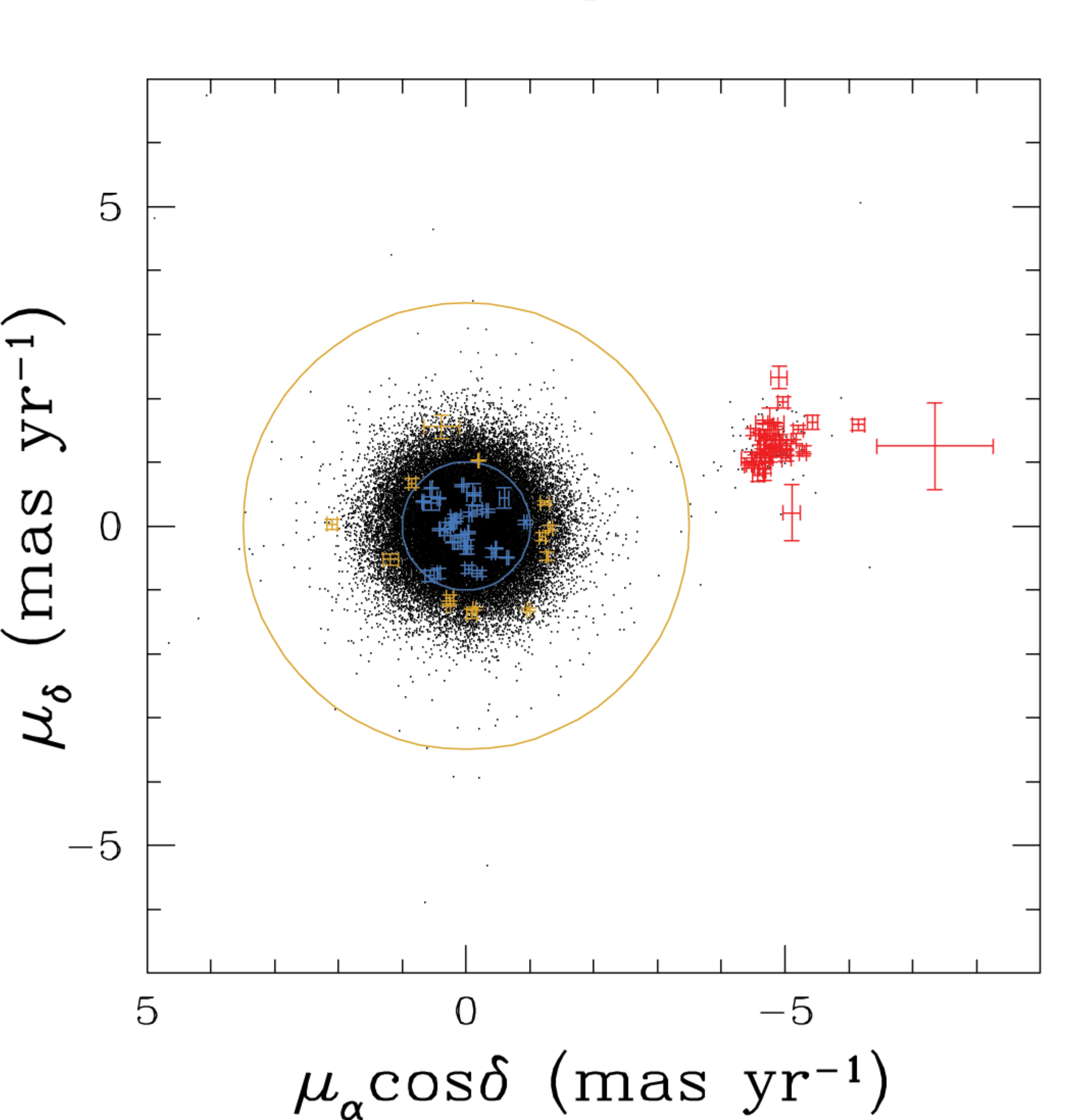}
\caption{Proper-motion diagram of all sources in \protect\cite{Bellini2014} catalogue (black). Sources that are most likely 47\,Tuc members are within the blue circle and marked in blue, while possible cluster members are in the ring area between the blue and yellow circles and are marked in yellow.  Most likely Small Magellanic Cloud (SMC) members are marked in red. The uncertainties in the proper-motion measurements are included for all coloured sources (blue, yellow and red).The radius of the blue and yellow large circles are defined as 1.5 and 5 times the field-averaged internal dispersion of cluster members. 
}
 \label{fig2}
\end{figure}

Thus, we obtained the information on the cluster membership of those objects by relying on the high-precision internal proper-motions obtained by \cite[][]{Bellini2014}. For the globular clusters included in the proper motion analysis from \cite{Bellini2014}, even in the most crowded central regions, accurate internal proper-motions of sources were measured. The methods used to determine the internal proper motions are able to reach a precision of $\sim$0.01 pixel per single exposure, for bright and unsaturated sources \citep[e.g.,][]{Anderson2006,Bellini2009,Bellini2011}. That is about 10 times smaller than the typical dispersion observed in a globular cluster 5\,kpc from the Sun over a 10-year time baseline, allowing the measurement of internal proper-motions of even rather faint sources.

We received high-precision internal proper-motion information obtained by \cite[][]{Bellini2014} from 73 of the sources bluer than the main-sequence and fainter than F275W=17 in the UV CMD (Fig. \ref{fig1}). Sources that do not have information on proper motion in the \cite{Bellini2014} catalogue were not included in our sample. However, that does not indicate that they do not belong to the cluster because those sources might be simply outside the field of view of the data to which the internal proper-motions have been determined. Among the 73 sources with proper-motion information, 44 are most likely members of the  SMC, 20 are most likely 47\,Tuc members and 10 are possible cluster members, with the latter still unlikely to not belong to the cluster (Fig. \ref{fig2}).

We found that 30 outliers that have internal proper motion consistent with the one of 47\,Tuc are bluer than the main-sequence and fainter than F275W=17 in the UV CMD (upper panel, Fig. \ref{fig1}). But, in the lower panel of Fig. \ref{fig1}, 24 of those are shown at the main-sequence in the F606W\,vs.\,F606W-F814W CMD (from now on referred as ``visual CMD'') or present significant shift towards the redder colours, i.e., they shift colours (``shifters''), while 6 outliers bluer than the main-sequence in both colour-magnitude diagrams, i.e., ``non-shifters'' (Fig. \ref{fig1}).

\subsection{Comparing with previous catalogues}
\label{previouscat}

\begin{table*}
\begin{tabular}{rlllllllccc}
\hline
ID & ID$_{Saraj.}$ & RA (\degr) & DEC (\degr) & F275W & F336W & F606W & F814W & PM & Shifter? & ID$_{BH17}$ \\ \hline
No & previous & classification  &   &   &   &   &   &   & \\ \hline
1  & 92909  & 6.050559  & -72.074784  & 19.979  & 19.092  & 18.382  & 17.833  & 1  & Yes & --- \\
2  & 14472  & 6.049497  & -72.105949  & 19.647  & 19.100  & 18.433  & 17.867  & 1  & Yes & --- \\
3  & 11483  & 6.065675  & -72.106537  & 19.847  & 19.174  & 18.489  & 17.912  & 1  & Yes & --- \\
4  & 94291  & 6.046181  & -72.077018  & 21.190  & 20.124  & 19.063  & 18.409  & 2  & Yes & --- \\
5  & 46584  & 6.060451  & -72.082201  & 21.491  & 20.364  & 19.093  & 18.456  & 1  & Yes & --- \\
6  & 16723  & 6.038436  & -72.099588  & 21.794  & 20.517  & 19.293  & 18.669  & 2  & Yes & --- \\
7  & 57032  & 6.026157  & -72.082666  & 22.287  & 21.035  & 19.381  & 18.517  & 1  & Yes & 371 \\
8  & 46593  & 6.060576  & -72.081903  & 19.349  & 19.416  & 19.622  & 19.603  & 2  & No & --- \\
9  & 48420  & 6.053474  & -72.081639  & 23.142  & 21.567  & 19.661  & 18.849  & 2  & Yes & --- \\
10  & 94778  & 6.044866  & -72.071966  & 22.393  & 21.124  & 19.689  & 18.873  & 1  & Yes & 254 \\
11  & 52899  & 6.038212  & -72.090284  & 22.699  & 21.397  & 19.724  & 18.999  & 2  & Yes & --- \\
12  & 43214  & 6.074299  & -72.090896  & 19.764  & 19.712  & 19.747  & 19.685  & 1  & No & --- \\
13  & 82503  & 6.101332  & -72.072490  & 22.851  & 21.420  & 19.761  & 19.084  & 2  & Yes & --- \\
14  & 53797  & 6.035949  & -72.085256  & 21.885  & 21.233  & 19.764  & 18.986  & 1  & Yes & --- \\
15  & 88579  & 6.067033  & -72.076294  & 23.327  & 21.784  & 19.954  & 19.182  & 1  & Yes & --- \\
16  & 93988  & 6.046538  & -72.075292  & 23.449  & 22.305  & 20.435  & 19.441  & 1  & Yes & --- \\
17  & 87267  & 6.072386  & -72.080282  & 24.237  & 22.589  & 20.441  & 19.501  & 1  & Yes & --- \\
18  & 39363  & 6.093210  & -72.081938  & 24.813  & 23.811  & 21.373  & 20.184  & 2  & Yes & --- \\
19  & 90383  & 6.059829  & -72.078171  & 23.717  & 23.889  & 21.458  & 20.387  & 2  & Yes & --- \\
20  & 87338  & 6.072945  & -72.076522  & 23.891  & 23.873  & 21.555  & 20.350  & 2  & Yes & --- \\
21  & 42265  & 6.078552  & -72.083123  & 23.082  & 23.208  & 21.713  & 20.530  & 2  & Yes & --- \\
22  & 43096  & 6.075476  & -72.083584  & 25.600  & 24.073  & 21.820  & 20.485  & 1  & Yes & --- \\
23  & 89174  & 6.064842  & -72.072471  & 25.383  & 24.368  & 22.227  & 20.858  & 1  & Yes & --- \\
24  & 15175  & 6.045630  & -72.100054  & 24.867  & 24.701  & 22.333  & 20.615  & 1  & Yes & --- \\ \hline
Some & previous & classification  &   &   &   &   &   &   & \\ \hline
25  & 57083  & 6.027156  & -72.081209  & 17.925  & 17.836  & 17.709  & 17.619  & 1  & No & 75 \\
26  & 46102  & 6.062164  & -72.094574  & 18.731  & 18.649  & 18.607  & 18.488  & 1  & No & --- \\
27  & 57456  & 6.025077  & -72.082279  & 19.053  & 19.579  & 20.472  & 19.482  & 1  & Yes & 30 \\
28  & 54823  & 6.032426  & -72.090924  & 21.438  & 20.903  & 20.251  & 19.652  & 1  & No & 21 \\
29  & 22861  & 6.008932  & -72.095020  & 19.077  & 19.360  & 20.452  & 19.870  & 1  & Yes & 56 \\
30  & 50798  & 6.044434  & -72.088140  & 22.024  & 21.312  & 20.858  & 20.591  & 1  & No & 71 \\
 \hline
\end{tabular}
\caption{List of the outliers detected by this work. Sources 1--24 have no previous classification, while outliers 25,26,27,28,29 and 30 were listed and classified in previous publications, mostly as CVs, with exception of outlier 26 that was classified as a W\,UMa (\protect\citealt{Albrow2001},\protect\citealt{Edmonds2003}, \protect\citealt{Knigge2008}, \protect\citealt{Sandoval2018}). Outliers 7 and 10 have X-ray sources within less than 0.5\,arcsec at \protect\citealt{Bhattacharya2017}'s catalogue (BH17).  PM=1 means the sources are most likely 47\,Tuc members, while PM=2 indicate possible cluster members.}
 \label{allout} 
\end{table*}

Our next step was to investigate the 30 objects that are, at least, likely cluster members in previously published catalogues. We compared the coordinates of our objects with the catalogues that aimed to search for binary stars, CVs, active binaries, millisecond pulsars, and radio sources \citep{McConnell2000, Albrow2001, Edmonds2003, Knigge2008,Miller-Jones2015, Ridolfi2016, Freire2017, Sandoval2018}. We also compared with the catalogue of variable stars in GCs from \citet[][and references there in, version 2017]{Clement2001}. Six outliers we detected in 47 Tuc had previous classifications summarized at Table \ref{previous}, while 22 of the shifters and 2 non-shifters we uncovered were not previously classified in any catalogue. 

 47\,Tuc is the GC with the most detected X-ray sources, with its most recent X-ray source catalogue being published by \cite{Bhattacharya2017}. They combined Chandra ACIS observations from 2000, 2002, and 2014--15 and identified 370 X-ray sources within the half mass radius of 47\,Tuc, reaching a luminosity in the 0.5--6 keV band of 1.0$\times10^{29}\,$erg\,s$^{-1}$. Knowing that the mean positional error of the X-ray sources detected by \citet{Bhattacharya2017} is $\sim$0.4\,arcsec, we compared the coordinates of our outliers with the ones from their table\,2 and found that there are 7 objects in our list that are within $\sim$0.4\,arcsec of their X-ray sources. Five of them have some previous classification, while 2 are not previously classified in any catalogue, and both are shifters (Table \ref{allout}). The clear majority of the sources we detected were not previously classified and have no X-ray source detected within 3 times the mean positional error of the X-ray sources: 20 shifters and 3 non-shifters listed in Table \ref{allout}. 
 
 \begin{table}
\singlespace
\center
\begin{tabular}{rclc}
  \hline
ID & ID$_{Saraj.}$ & Previous Classifications & ID$_{BH17}$\\ 
\hline
25 & 57083 & SPV?$^{1}$, AB$^{2}$, Binary$^{4}$, CV?$^5$ & 75 \\
26 & 46102 & W UMa $^{1}$ & --- \\
27 & 57456 & Variable$^1$, CV?$^{2}$, CV (WD--MS)$^{4}$, CV$^5$ & 30\\
28 & 54823 & CV?$^{2}$, Intriguing object$^{4}$ & 21 \\
29 & 22861 & CV?$^{2}$, CV$^{5}$ & 56 \\
30 & 50798 & CV?$^{2}$,  CV?$^{3}$  &  71 \\
\hline
 \end{tabular}
\caption{The Sarajedini's ID of outliers we detected in 47\,Tuc that were listed in previous publications. All sources are most likely cluster members according to \protect\citealt{Bellini2014}. $^{1}$\protect\citealt{Albrow2001}, $^{2}$\protect\citealt{Edmonds2003}, $^{3}$\protect\citealt{Heinke2005},  $^{4}$\protect\citealt{Knigge2008}, $^{5}$\protect\citealt{Sandoval2018}. AB= active binaries; BY-Dra= BY Draconis variables; CV= Cataclysmic variable; CV?= Cataclysmic variable candidate; SPV= Short Period Variable; SG= Sub Giant; V= Variable; W UMa= W Ursae Majoris variable; WD--MS= White dwarf--Main-sequence binary. X-ray ID by \protect\citealt{Bhattacharya2017} (BH17). }
 \label{previous} 
\end{table}
 
\subsection{Blending and Contamination}

A question that arises is, could outliers be the result of the superposition of two or more unrelated stars from the cluster that are far from each other but along the same line of sight? The argument against this idea comes from the internal proper-motions obtained by \cite{Bellini2014}. The range of data used to determine the proper motion of sources near the centre of 47\,Tuc was 10\,years considering a wide range of HST filters. In 10 years, 47\,Tuc stars move on average by about half an ACS pixel (i.e., ~25\,mas) with respect to each other, so that a blend of unrelated sources would change the shape of the sources dramatically from the first to the last epoch, producing large proper-motion errors, especially if one source is blue and the other is red, and images of different epochs were taken using different filters. However, the uncertainties of the proper-motion of our outliers are small and consistent with being single sources or unresolved binaries. 

We also evaluated the probability of field contamination of objects with similar proper motions as 47\,Tuc using the Besan\c{c}on models \citep{Robin2003}. We ran the simulation for a 1 square degree field towards 47\,Tuc. It returned 8755 stars brighter than V=23, including their magnitudes and proper motions. Then, we used Gaia DR2 data \citep{Helmi2018} to get the mean proper motion of 47\,Tuc($\mu_{\alpha}$=5.2\,mas/yr, $\mu_{\delta}$=-2.5\,mas/yr) and estimated the proper motion dispersion as being 0.59\,mas/yr, by using the results present on \cite{Kamann2017}. 

We found that, if we only count stars with proper motions within 1.5 times the proper motion dispersion of 47\,Tuc, we get 309 stars per square degree (or 0.09\,stars/square arcmin). If we consider 5 times the proper motion dispersion, the number of stars per square degree is 1398 (or 0.39\,stars/square arcmin). The visual data we are using was obtained with ACS which has a field of view of 11.36 square arcmin. The WFC3 UV data that superposes over the visual data only covers half of the ACS area, 5.68 square arcmin. Combining those numbers with the results from the Besan\c{c}on models we get that there can be 0.51 or 1.56\,stars from the Galactic field with proper motions within 1.5 and 4 times the proper motion dispersion of 47\,Tuc. Therefore, it is unlikely that our outliers are objects from the Galactic field with the same proper motion of 47\,Tuc. Contamination from background SMC stars is even less probable, given the clear proper-motion separation between 47 Tuc and SMC stars (see Fig. \ref{fig2}).

The last question in this point is: could the behaviour of the shifters be explained by them being faint blue objects contaminated by red bright stars or diffraction spikes close by in the redder filters? To check if that is the case for our shifters, we inspected each of our outliers in F814W and F336W images (see the finding charts of each of our 24 outliers in Appendix \ref{Appendix A}). From that examination we were able to show that the outliers are clearly individual sources, with no contaminations from nearby sources. Hence, we conclude that the behaviour presented by the shifters is not caused by photometric contamination by nearby bright or saturated red stars in redder filters. 

\subsection{The Observed Colour-Magnitude Diagram}
\label{cmdsection}

Aiming for the better visualisation of the position of the stars that follow single stellar-evolution in the CMD, we over plotted isochrone models of 11.0\,Gyr \citep{Campos2016} and \textit{Z}=0.003 \citep[][2010 edition]{Harris1996} in all our CMDs of 47\,Tuc data. For that we combined models by {\sc mesa}\footnote{\url{http://mesa.sourceforge.net/}} Isochrones and Stellar Tracks \citep[MIST, ][]{Dotter2016, Choi2016} with the ones from Bag of Stellar Tracks and Isochrones \citep[BaSTI,][]{Salaris2010}. The reason for combining two sets of models is the fact that the MIST models only cover from the main-sequence down to white dwarfs with ${T}_\text{eff}\sim$23\,000\,K. So, because we wanted to show the isochrone models from the low mass main-sequence stars down to the WD cooling sequence, we had to add the BaSTI models for WDs cooler than ${T}_\text{eff}\sim$23\,000\,K.    

The first step toward understanding those outliers that have no previous classification is to investigate the possible known explanations. One possibility is that the behaviour of those sources could be explained by variability induced by pulsation, rotation or eruption \citep[e.g.][]{Eyer2018}.  However, as mentioned in Section \ref{phot}, the magnitudes were determined by averaging the values obtained in several images for the same filter in multiple epochs, and the photometric scatter was set as the uncertainty in the magnitude of a particular star. It can be seen that the scatter  in the magnitudes in the visual CMD are very small, while in the UV CMD the uncertainties for outliers fainter than F275W=24 are higher, but not enough to explain their behaviour as being caused by variability induced by pulsation, rotation or eruption.

The shifters that do not present any detectable X-ray source nearby cannot be explained by the most up-to-date single evolution stellar tracks. However, they could be explained as binary systems composed of a WD and a main-sequence star (WD--MS). Such systems could be primordial, i. e., established just after the gas has been removed from the forming system \citep{Kouwenhoven2005}, or be more recently formed due to the crowded environment of globular clusters. Additionally, the WDs in the WD--MS binaries could be normal-mass WDs \citep[M$\goa$0.53M\,$_{\odot}$,][]{Moehler2004,Kalirai2009} or WDs with masses lower than the ones expected from single stellar-evolution within the GC's age, indicating a transfer of mass due to binary stellar-evolution.  

\begin{figure}
\subfigure[\label{subfig3a}]{\includegraphics[width=0.17\textheight, trim={1cm 2.5cm 1cm 2.5cm},clip]{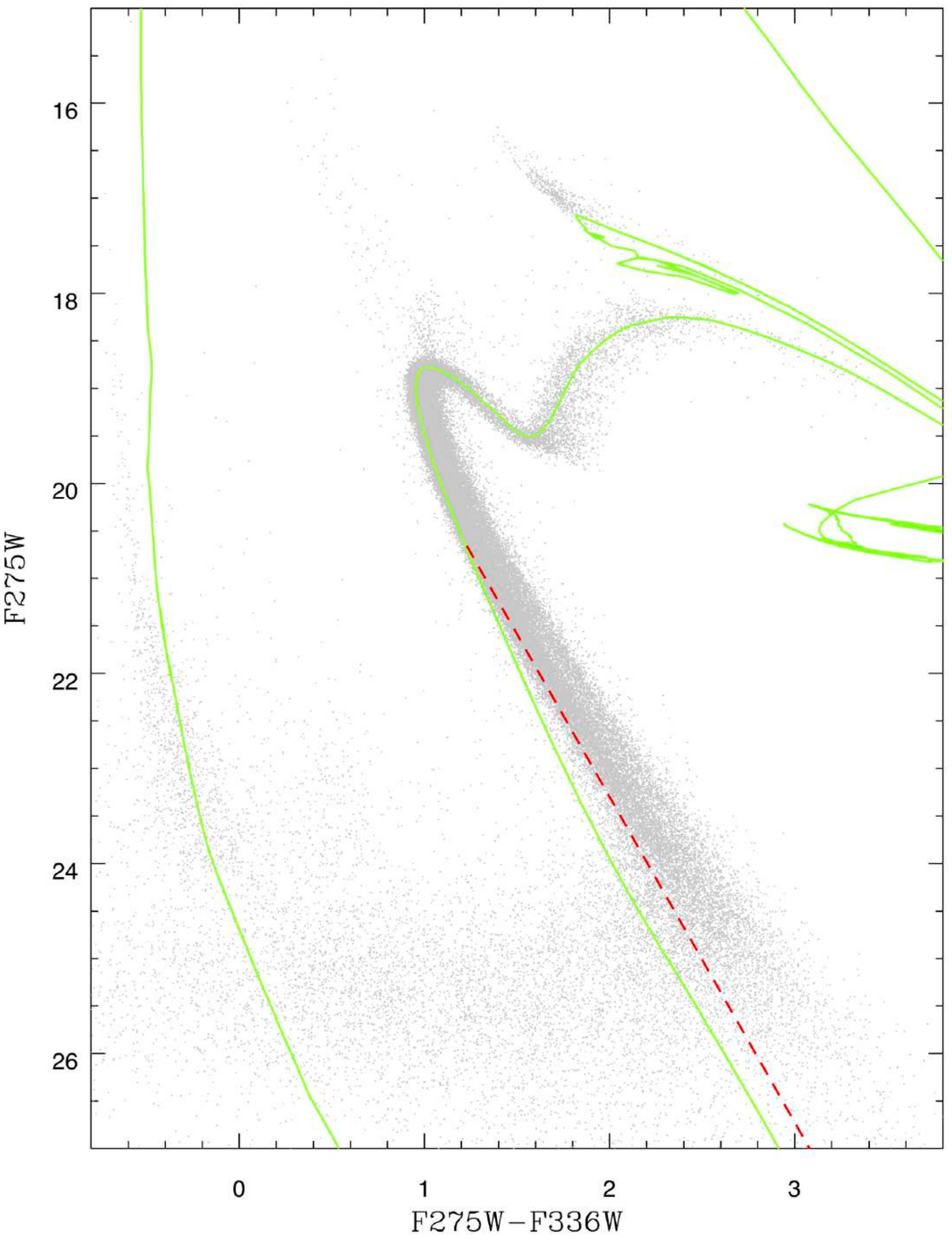}}
\subfigure[\label{subfig3b}]{\includegraphics[width=0.17\textheight, trim={1cm 2.5cm 1cm 2.5cm},clip]{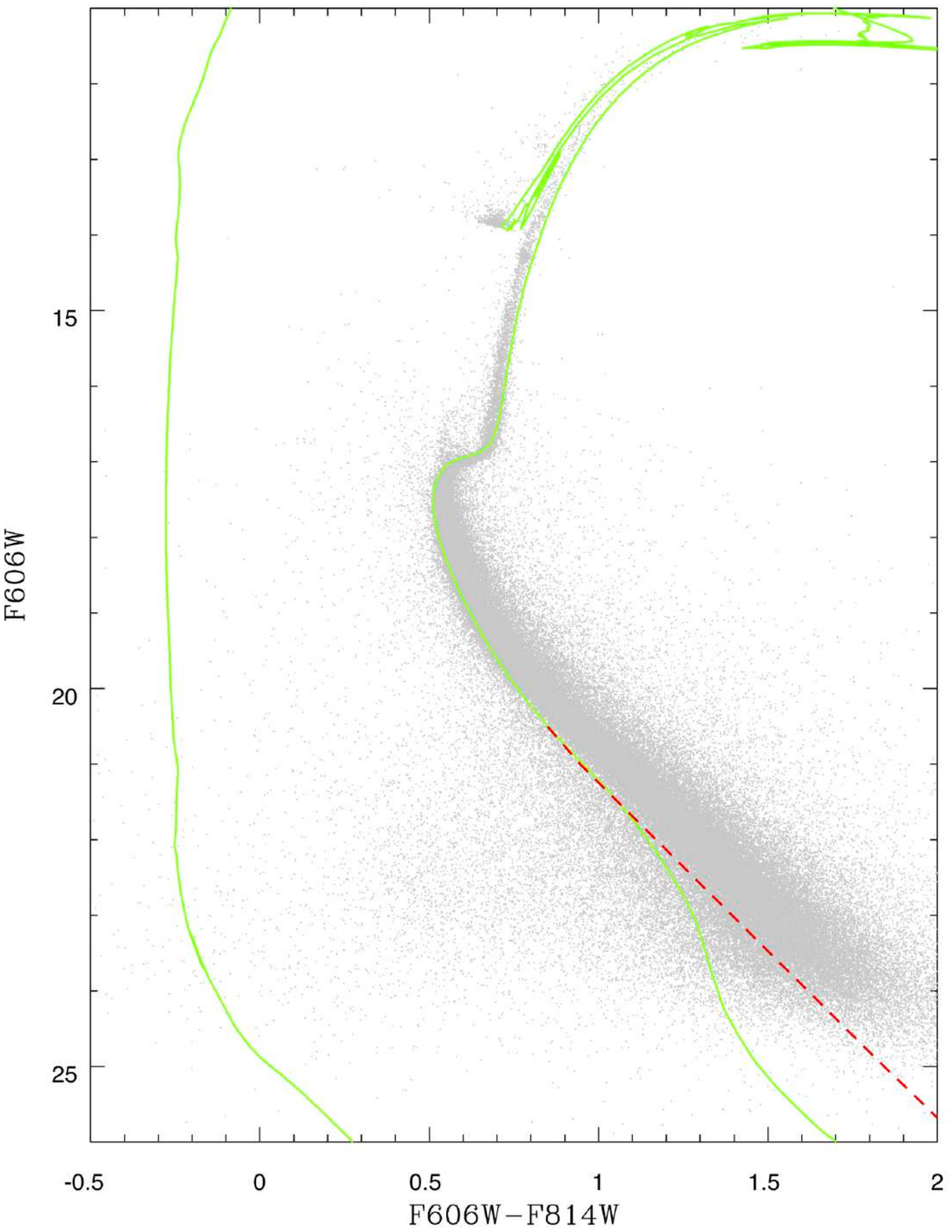}}
\caption{The colour-magnitude diagrams (dark grey) in F275W\,vs.\,F275W-F336W (panel a) and F606W\,vs.\,F606W-F814W (panel b). The green lines the same as the ones presented in Fig \ref{fig1}. The red dashed line denotes the correction applied to the low main-sequence models, in order for them to follow the slope of the observed data. }
\label{corrmodels}
\end{figure}

To explore the WD--MS possibility, we built our WD--MS models by combining the main-sequence portion of the isochrone with magnitudes of normal mass WDs and ELMs, obtained from \cite{Salaris2010} and \cite{Istrate2016}, respectively. 
A well-known problem is that models of very low mass stars tend to have optical colours which are far too blue at the low main-sequence \citep{Campos2013, Chen2014}. Such mismatch is evident in both CMDs presented in Fig. \ref{fig1}, where the isochrone models begin to deviate from the data in the lower main-sequence at F606W$\sim$21.5 for the visual colours (Figs. \ref{subfig1a}, \ref{subfig3a}) and F275W$\sim$20.5 for the UV colour (Figs. \ref{subfig1b}, \ref{subfig3b}). With that in mind, we applied an empirical  correction for the lower main-sequence portion of the models, shown in Fig. \ref{corrmodels}, so that the model follows the slope of the data.

\begin{figure*}
\subfigure[\label{subfig4a}]{\includegraphics[width=0.33\textheight, trim={1cm 1.5cm 1cm 1.8cm},clip]{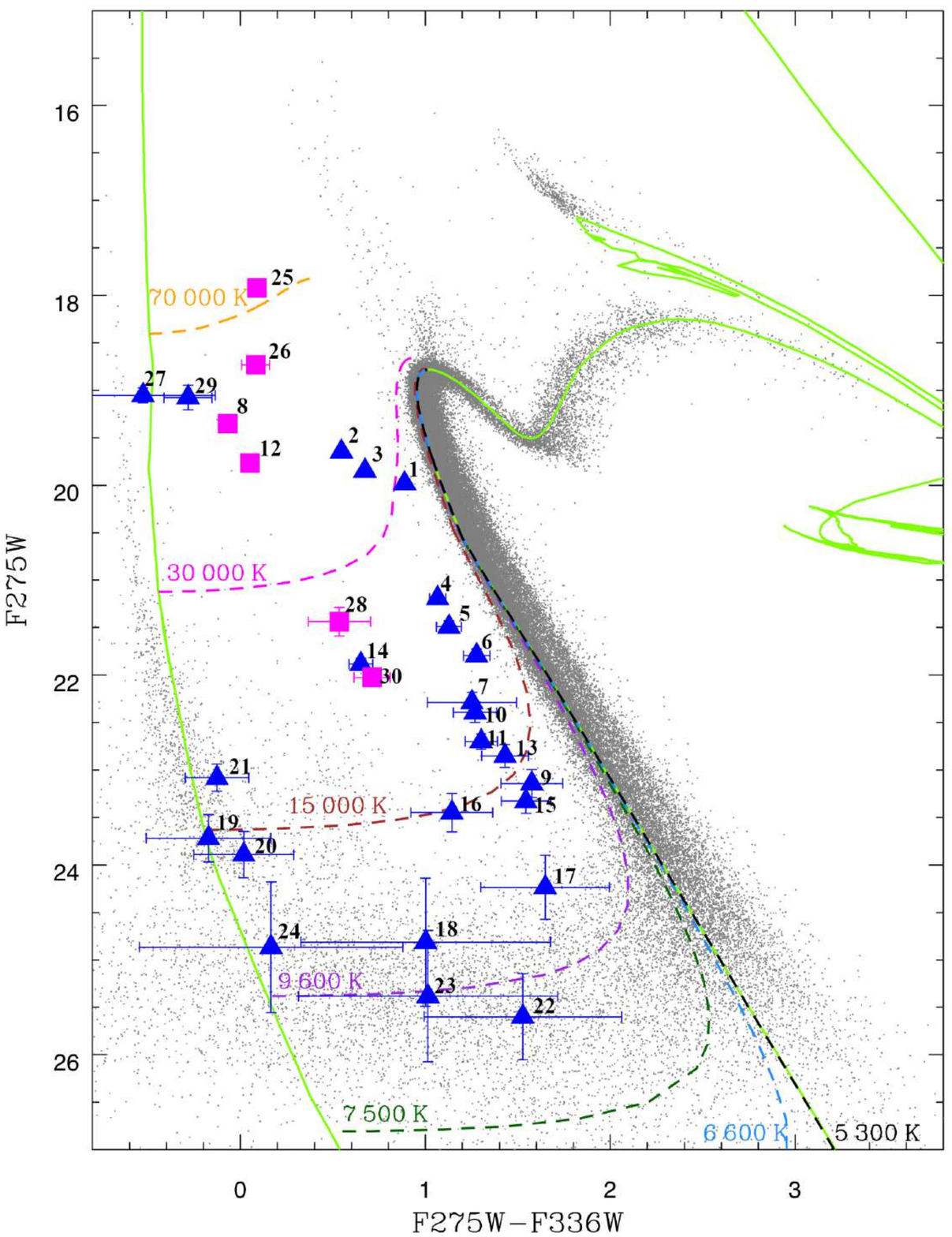}}
\subfigure[\label{subfig4b}]{\includegraphics[width=0.33\textheight, trim={1cm 1.5cm 1cm 1.8cm},clip]{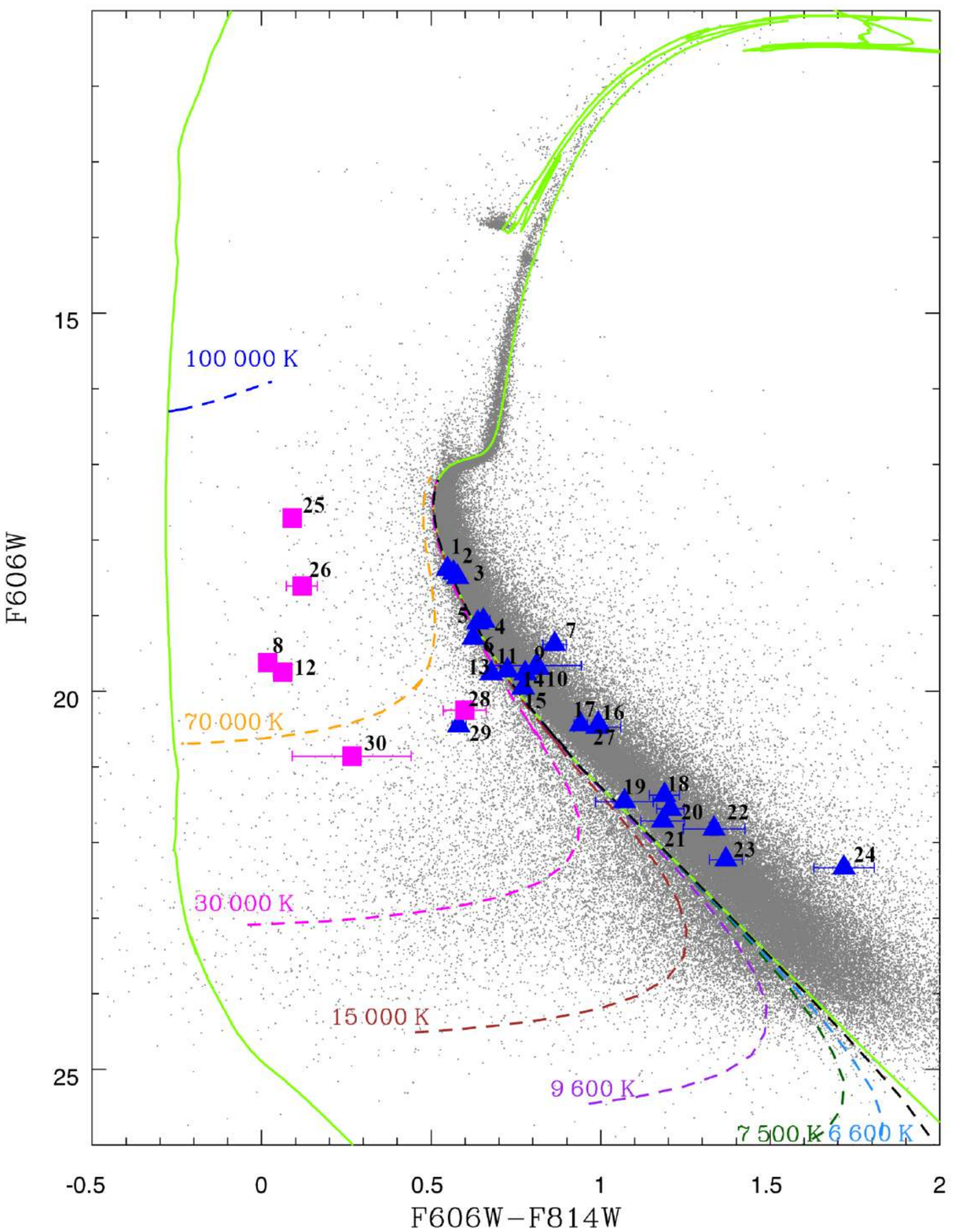}}
\subfigure[\label{subfig4c}]{\includegraphics[width=0.33\textheight, trim={1cm 1.5cm 1cm 1.8cm},clip]{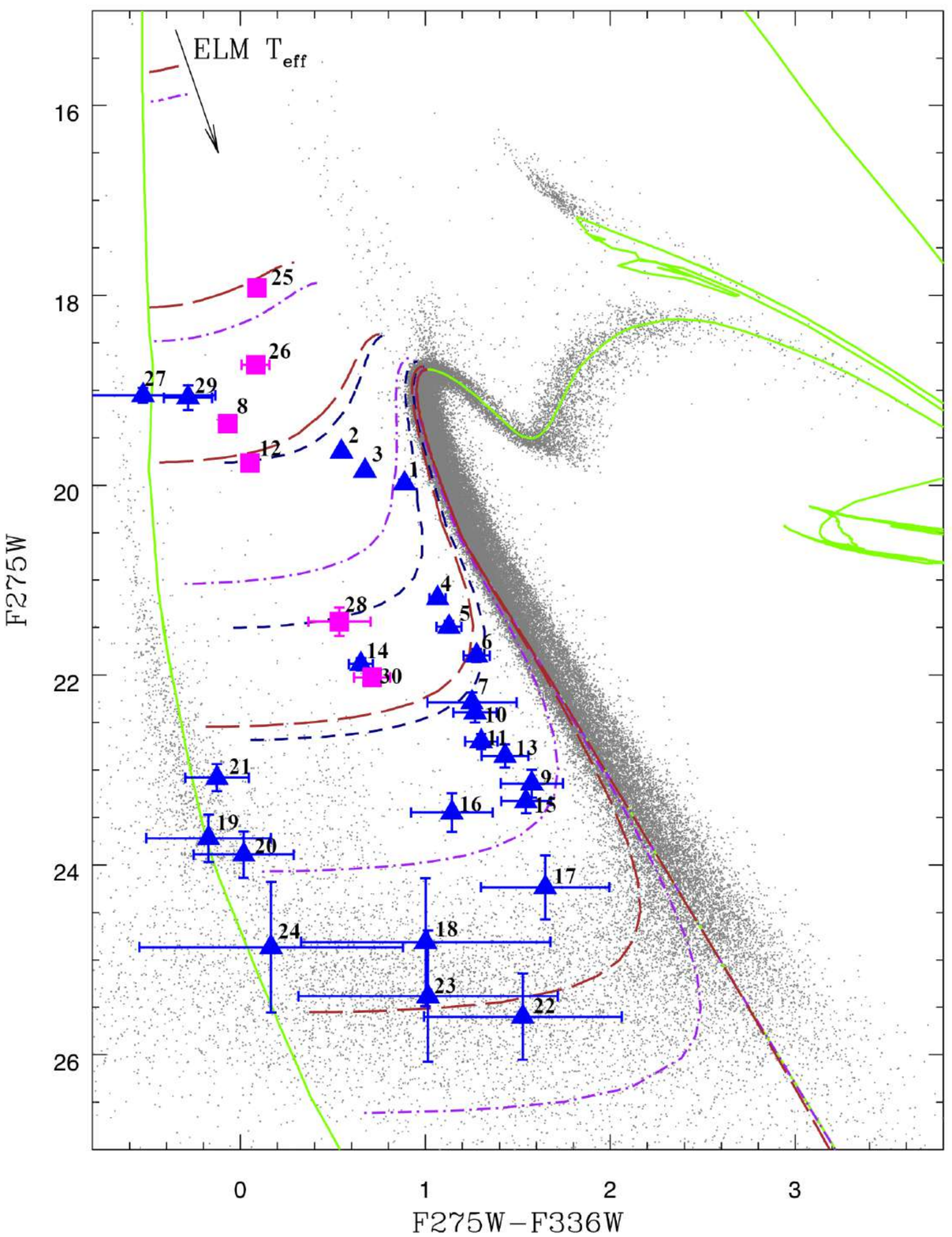}}
\subfigure[\label{subfig4d}]{\includegraphics[width=0.33\textheight, trim={1cm 1.5cm 1cm 1.8cm},clip]{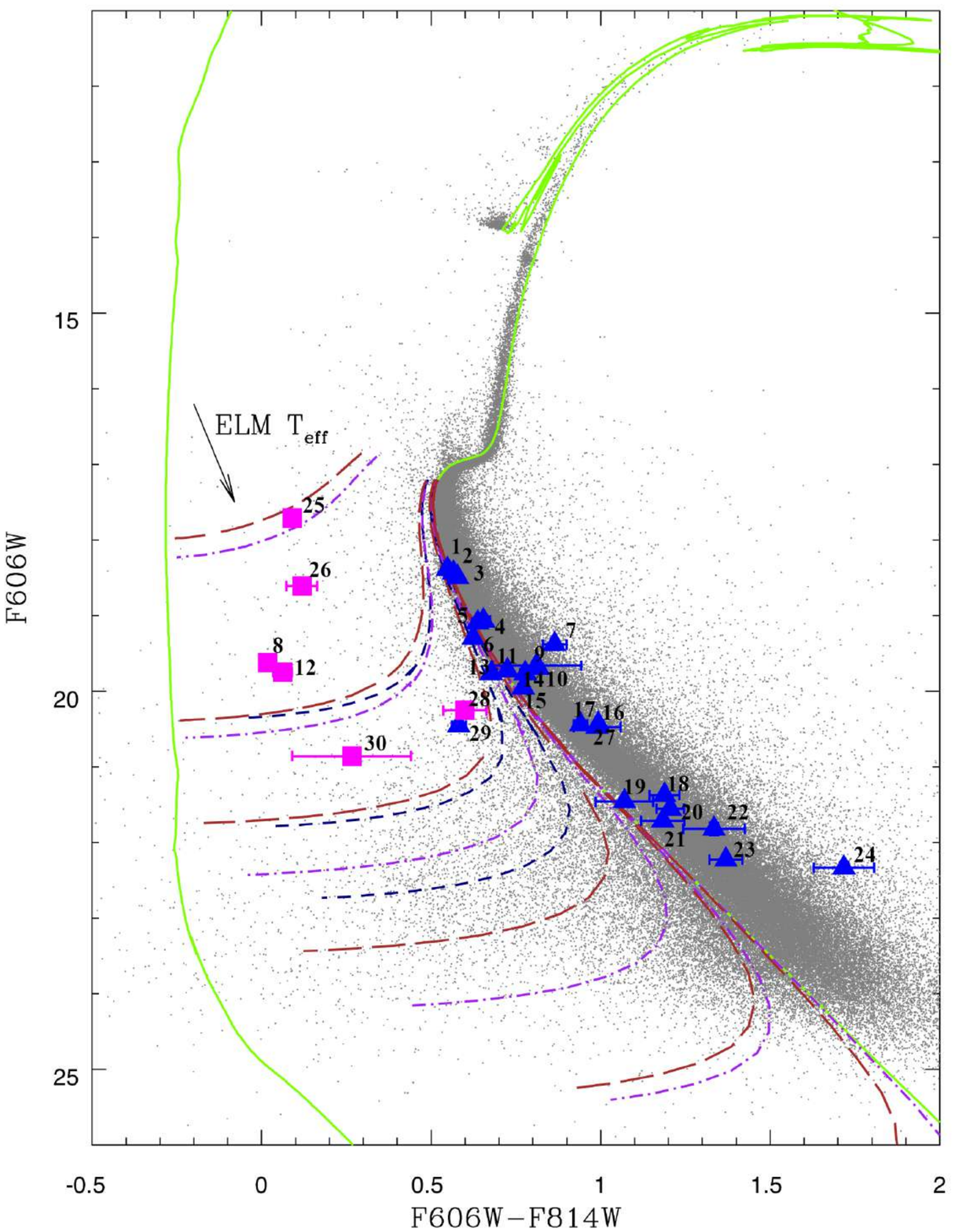}}
\caption{The colour-magnitude diagrams (dark grey) in UV (panels a and c) and visual (panels b and d). All the outliers that are most likely cluster members are marked as light-blue triangles and possible cluster members are marked as yellow squares. The green solid lines are MIST \protect\citep{Dotter2016,Choi2016}, with the correction for low main-sequence, and BaSTI \protect\citep{Salaris2010} isochrones with 11.0\,Gyr and \textit{Z}=0.003. In panels a and b, the dashed lines represent models of WD+MS composed by normal-mass WDs and main-sequence stars. In panels c and d, models of WD+MS composed by ELM WDs and main-sequence stars are shown, with ${T}_\text{eff}\sim$ decreasing from top to bottom (140\,000\,K to 5\,000\,K). The masses of the ELMs are 0.18\,M$_{\odot}$ (short-dashed line), 0.25\,M$_{\odot}$ (dot-dashed line), 0.32\,M$_{\odot}$ (long-dashed line).}
 \label{fig4}
\end{figure*}

The first result we notice from comparing the WD--MS models composed of normal mass WDs with the ones composed of ELMs is that they populate the exact same area of the CMD, and their behaviour, when comparing the UV and visual CMDs, is very similar to the one presented by the shifters we found (Fig. \ref{fig4}). That enforces the possibility that the shifters are WD--MS binaries. However, it does not provide the answer to whether  they might be composed of a main-sequence star plus a normal mass WD or an ELM. 

From Fig. \ref{fig4} we can also rule out WD--MS as an explanation for our non-shifters (8 and 12). When looking at the temperature of the models, for normal-mass WDs the position of sources 8 and 12 in the UV CMD is consistent with WD--MS models with 30\,000\,K <${T}_\text{eff}$<70\,000\,K, but in the visual CMD they are between 70\,000\,K <${T}_\text{eff}$<100\,000\,K. While for ELMs, sources 8 and 12 are located between WD--MS models with 30\,000\,K <${T}_\text{eff}$<50\,000\,K  in the UV CMD and  50\,000\,K <${T}_\text{eff}$<70\,000\,K in the visual CMD. Outlier\,18 is consistent with WD--MS models with 15\,000\,K <${T}_\text{eff}$<30\,000\,K in the UV CMD, but between 30\,000\,K <${T}_\text{eff}$<70\,000\,K in the visual CMD. While for ELMs, source\,18 is between WD--MS models with 15\,000\,K <${T}_\text{eff}$<30\,000\,K  in the UV CMD, and  30\,000\,K <${T}_\text{eff}$<50\,000\,K in the visual CMD.

\begin{figure}
\subfigure[\label{subfig5a}]{\includegraphics[width=0.32\textheight, trim={1cm 1.5cm 1cm 1.8cm},clip]{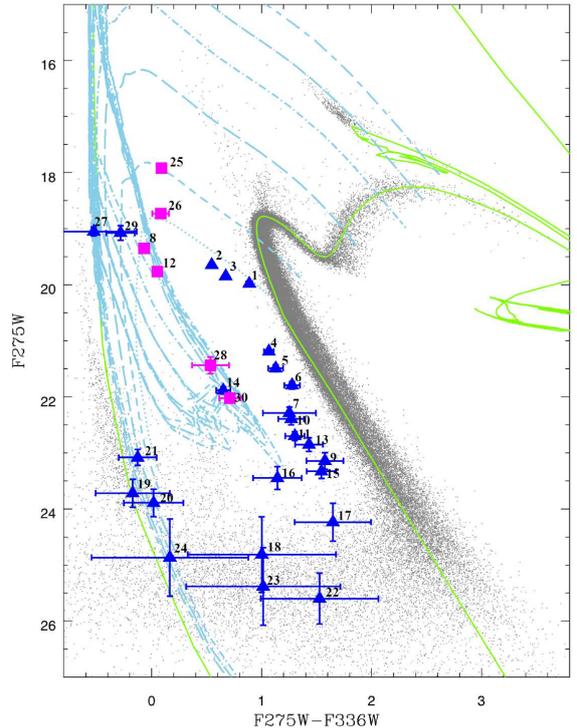}}
\subfigure[\label{subfig5b}]{\includegraphics[width=0.32\textheight, trim={1cm 1.5cm 1cm 1.8cm},clip]{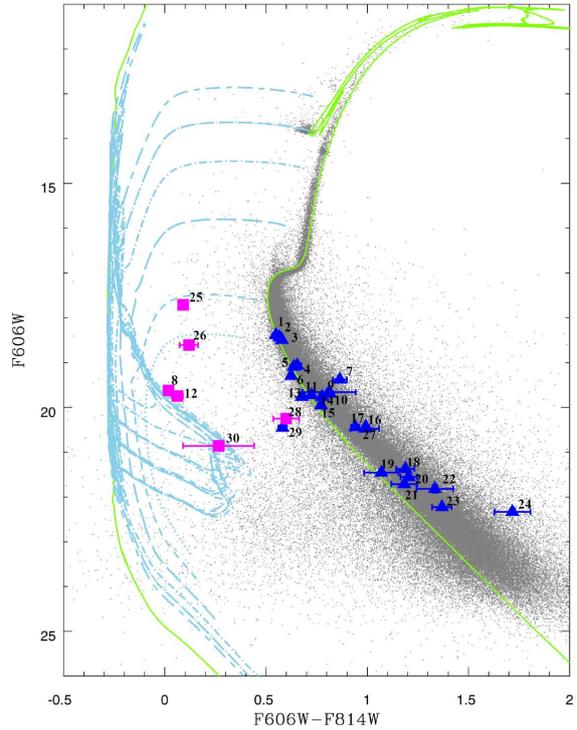}}
\caption{The CMDs (dark grey) in F275W\,vs.\,F275W-F336W (panel a) and F606W\,vs.\,F606W-F814W (panel b). The shifter outliers are marked as blue triangles, and the non-shifters are marked as magenta squares. We included the models for ELMs, with \textit{Z}=0.001, during the proto-WD phase and WD cooling phase for the 0.18\,M$_{\odot}$ (dot line), 0.20\,M$_{\odot}$ (short dash line), 0.23\,M$_{\odot}$ (long dash line), 0.26\,M$_{\odot}$ (short dash-dot line), 0.30\,M$_{\odot}$ (long dash-dot line), and 0.32\,M$_{\odot}$ (long dash-short dash line). The solid green line is composed of MIST \protect\citep{Dotter2016,Choi2016} and BaSTI \protect\citep{Salaris2010} isochrones with 11.0\,Gyr and Z=0.003.}
\label{fig5}
\end{figure}

One possible explanation for the non-shifters can be that they are ELMs or pre-ELMs that may or may not be double-degenerate systems. It is well-known that white dwarfs with lower masses appear redder than the more massive ones in the CMD \citep[e.g., ][]{Fontaine2001,Bellini2013}. The reddest white dwarfs that might appear in the gap between the WD cooling sequence and the main-sequence in the CMD of globular clusters are the ELMs. Whether those ELMs are found as companions of massive degenerate objects (neutron star, WD) or they are detected as single stars, they would predominantly appear as blue objects in the CMD, even when using multiple colours (i.e., they would mostly be in the gap between the main-sequence and the WD cooling sequence in both visual and UV CMDs). We show the CMD of our data along with evolutionary models for ELMs from \citep{Istrate2016} formed by stable Roche-lobe overflow mass transfer in a LMXB during the proto-WD phase and WD cooling phase. The range of final masses for the ELMs showed in Fig. \ref{fig5} covers masses from 0.18--0.32\,M$_{\odot}$ and Z=0.001, the computed metallicity closest to the one of 47\,Tuc. 

A possible explanation for the non-shifter outliers 8 and 12 (highlighted in Fig. \ref{fig5}) arise when comparing them to the ELM and pre-ELM models from \cite{Istrate2016}. The  sources 8 and 12 can be ELMs in the proto-WD phase, undergoing hydrogen shell flashes, because their position in both CMDs is precisely where the models predict the shell flashes to occur. There are no models for masses lower than 0.18\,M$_{\odot}$ with \textit{Z}=0.001 \citep{Istrate2016} due to the constraint imposed by the bifurcation period that limits the minimum mass for an ELM to be created from binary evolution by stable Roche-lobe overflow mass transfer \citep{Istrate2014}. Of course, ELMs less massive than 0.18\,M$_{\odot}$  may be produced in systems without neutron star companions, as they apparently are in the Galactic disk \citep{Brown2016}.

Sources\,8 and 12 could be possible ELMs with masses lower than 0.18\,M$_{\odot}$, since ELMs with masses lower than 0.18\,M$_{\odot}$ would populate an area slightly redder and fainter than the ones with higher masses. if that possibility is confirmed it would demonstrate that our level of understanding on the multiple channels of formation of ELMs is still incomplete. 

\subsubsection{Outlier 26}
\label{46102}

\begin{hyphenrules}{british}
\hyphenation{bi-na-ries}
\end{hyphenrules}

Source\,26 has been previously classified as a possible W UMa binary by \cite{Albrow2001}. However, \cite{Albrow2001} showed that this source does not obey the Rucinski period-luminosity-colour-metallicity relation for contact binaries, such as the W UMa binaries \citep{Rucinski1994, Rucinski1995}. As a matter of fact, from all the W UMa candidates from \cite{Albrow2001}, source\,26 is the one that deviated the most from the Rucinski calibration. Furthermore, the unequal eclipse maxima present in the light curve of source\,26 indicates that it might be a semi-detached system, incompatible with the W UMa binaries description.   

Those features made \cite{Albrow2001} attempt to find other explanations for outlier\,26. They argued that source\,26 could be a CV because it presents very blue colours that could indicate mass transfer from the secondary causing a hot spot on the surface of the primary or a surrounding accretion disk. However, we found no X-ray source near source\,26 on the \cite{Bhattacharya2017} catalogue, making the CV possibility less likely. 

One possibility that was not explored by \cite{Albrow2001} is that outlier\,26 could be a double-degenerate system where one of the stars is an ELM. Outlier\,26 is in the gap between the main-sequence and the WD cooling sequence both in UV and visual CMDs. When comparing to the ELMs and pre-ELMs models from \cite{Istrate2016}, source\,26 is very close to the $\sim0.18\,M_{\odot}$ model in both CMDs. 

Additionally, the light curve obtained by \cite{Albrow2001} for source\,26 indicates it has a variability period of 0.259\,days. Such period is consistent with the ones found for some double degenerate systems from Sloan Digital Sky Survey (SDSS) studied by \cite{Brown2016} as part of the ELM survey. Those stars have an average period of 0.255\,days, an average mass of $\sim0.182\,M_{\odot}$ for the ELM and an average mass of the companion of $\sim1.00\,M_{\odot}$. \citet{Strickler2009} inferred similar masses for unseen companions to the ELMs in NGC\,6397."All that makes the possibility of outlier\,26 being a double degenerate system, composed of an ELM and a massive WD, be extremely feasible. Confirmation might only be obtained with spectroscopic data. 

\subsection{H\texorpdfstring{$\alpha$}\,\,photometry  }
\label{Halpha}

\begin{figure}
\center
 \includegraphics[width=0.32\textheight, trim={1cm 2.5cm 1cm 3cm},clip]{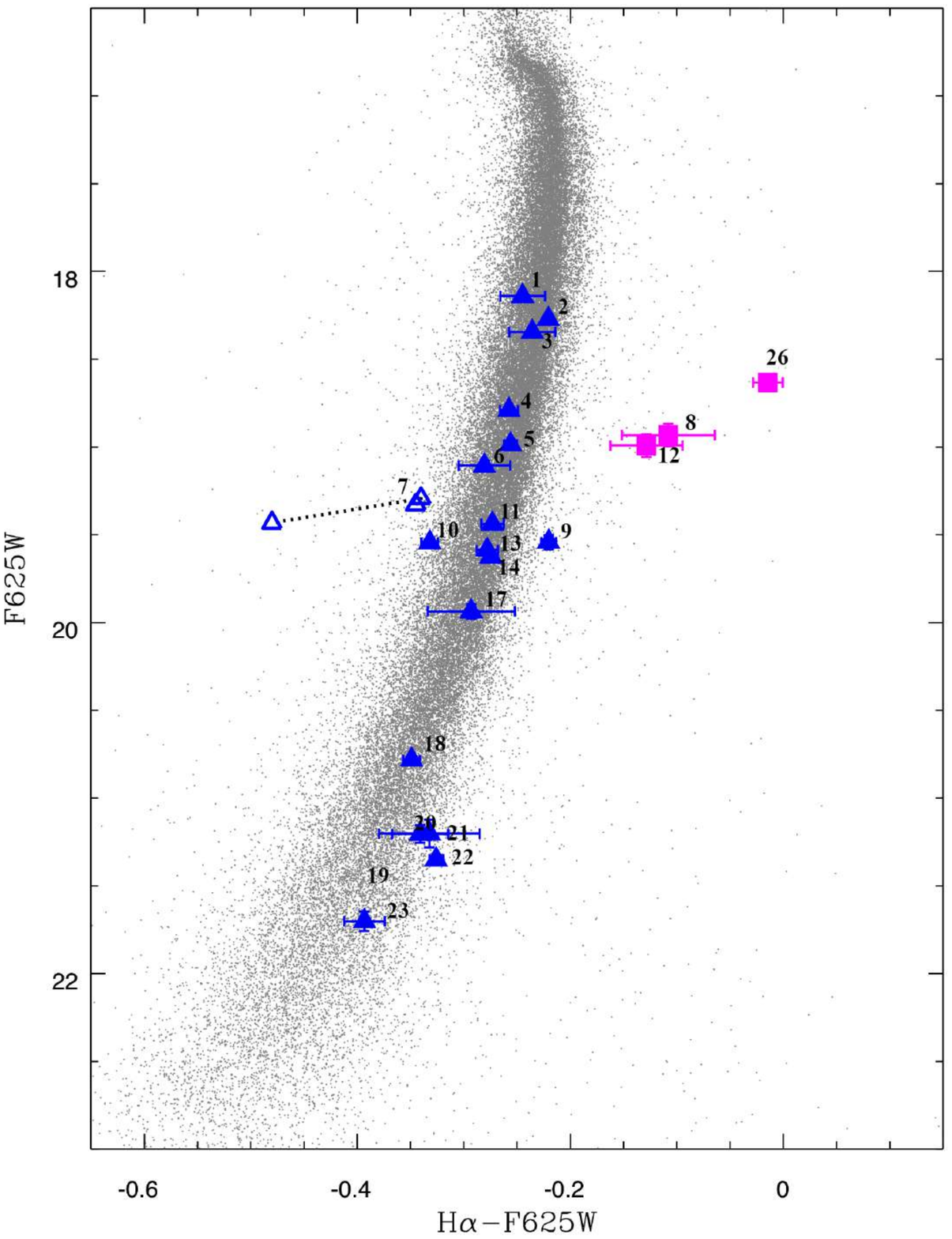}
 \caption{ F625W versus H$\alpha$--F625W CMD for 47\,Tuc stars built by averaging 3 epochs of data obtained in 2002. The shifter outliers are marked as blue triangles, and the non-shifters are marked as magenta squares. The error bars show the photometric dispersion of the three epochs. The position of the ELM candidates 8, 12 and 26  shows that they are consistent with being ELMs. Source 7, a counterpart to an X-ray source, presents clear variability within the 3 epochs, for that reason we show the individual photometry in each epoch for this source. In 2 of the epochs the source is very close to the main-sequence, while in 1 epoch it presents an emission in H$\alpha$, being consistent with the idea that this star is currently going through some kind of mass transfer. 
}
 \label{halpha}
\end{figure}

The F658N (H$\alpha$) band can help us to identify objects with large H$\alpha$ emission/absorption. With that in mind, we obtained the photometric data of 47\,Tuc available on the Hubble Legacy Archive (HLA) and were able to find photometry for 20 out of our 24 outliers with no previous classification, due to the fact of them being too faint or out of the field of view. We also found source\,26, the ELM candidate with a previous classification as a W UMa by \citet{Albrow2001}. 

In Fig. \ref{halpha} we present the F625W\,versus\,H$\alpha$-F625W CMD. This was constructed by averaging the photometry of the three different epochs of each filter. After that, we determined the RMS and set it as the uncertainty in the magnitude of each star. To avoid an excess of lines in the figure we only show the uncertainty bars for the outliers and source\,26. 

The first feature to be noticed is the position of sources 8, 12 and 26. These objects were already considered ELM candidates, based on our analysis of the visual and UV CMDs, and now they are all found to have significant H$\alpha$-absorption, as expected for ELMs. Making these objects even more consistent with the proposed classification as ELM candidates. 

Secondly, source 7 is a possible counterpart to the newly discovered X-ray source 371 of the \citet{Bhattacharya2017}'s catalogue. This is the only source for which we have included the individual photometry of each epoch instead of the averaged magnitude. That is because source 7 presents clear variability within the 3 epochs, in 2 of the epochs the source is very close to the main-sequence, while in 1 epoch it presents an emission in H$\alpha$. That feature shows that source 7 is currently going through some kind of mass transfer that seems to vary in intensity with time.

\section{Discussion}
\label{discussion}
The high stellar densities present in GCs cause close encounters between stars and binaries to be much more frequent than in the Galactic field. Those dynamical interactions can not only form new binaries but also harden, modify or destroy an existing binary system. The stars composing such systems can follow a single stellar-evolution path or become stellar exotica and evolve as a non-single star. 

Our analysis of multicolour photometry of 47\,Tuc revealed a population of evolved binary systems with at least one of the stars likely being a WD. Part of those systems could be the result of evolution of primordial binaries while some of them are likely dynamically formed \citep[e.g.][]{Ivanova2005}. Thirty-six of our outliers are shifters, sources that shift positions when observed in visual and UV CMDs, probably WD--MS systems, with 2 of them having X-ray sources within less than 0.4\,arcsec. Three are non-shifters, double degenerate systems candidates composed of a massive WD or a neutron star and an ELM or pre-ELM. 

Spectroscopic data for one of our shifters (source\,2) was found among the 19181 spectra obtained for 47\,Tuc by \citet{Kamann2017} with the Multi Unit Spectroscopic Explorer \citep[MUSE; ][]{Bacon2010}. Unfortunately, the individual spectra for the star had low signal-to-noise ($S/N \lesssim 5.0$) preventing the determination of reliable parameters through radial velocity analysis. However, the radial velocity spread presented in the data of outlier\,2  is unlikely to be caused by statistical uncertainty to a 90\% confidence level, suggesting that it might indeed be a binary system, as our photometric analysis suggests.

The first question that one might have is: why can't these binary systems be composed of a main-sequence and a neutron star? The answer comes from the comparison between quiescent X-ray luminosities of black hole X-ray novae and neutron star X-ray novae. For comparable orbital periods, the black hole X-ray novae are 2 to 3 orders of magnitude fainter than Neutron Star X-ray Novae \citep[e.g., ][]{Narayan1997, Garcia2001, Narayan2008}. As discussed in Section \ref{previouscat}, the faintest X-ray luminosity detected in 47\,Tuc is L$_X\sim$10$^{29}$\,erg\,s$^{-1}$ \citep{Bhattacharya2017}, while black hole X-ray novae present quiescent luminosities as low as L$_X\sim$10$^{30}$\,erg\,s$^{-1}$ \citep{Garcia2001}. So, if any of our sources were composed of a main-sequence star plus a neutron star, they should present observable nearby X-ray sources within the \cite{Bhattacharya2017} catalogue. 

There are various possible formation channels for WD--MS binaries. As mentioned earlier, they can be primordial main-sequence--main-sequence (MS--MS) binaries in which the more massive star evolved and became a WD. Another formation channel of the WD--MS is through dynamical interactions. The high densities present in GCs can form, disrupt, and even modify binary systems.

The second question that arises when looking at our sample is: is it possible that some of our WD--MS systems might be CVs? CVs can be detected in GCs through a number of methods: variability, emission lines, extremely blue colour and/or X-ray emission \citep{Knigge2011}. Most CVs and CV candidates in GCs have been identified by their X-ray emission \citep{Sandoval2018}. 47\,Tuc is the GC with the most CVs identified so far, with 43 CVs and CV candidates \citep{Sandoval2018}. However, this number is far lower than the number of CVs expected to be found for a typical massive cluster \cite[$\sim$200, according to][]{Ivanova2006}. We have no information about the variability of our outliers.  However, all of our outliers present extremely blue colour. 

When compared with the \cite{Bhattacharya2017} catalogue, only 2 of our outliers present an X-ray source within 0.4\,arcsec, the mean positional error of the X-ray sources, which, along with those presenting UV excess, allow us to classify them as CV candidates. In addition to that, source\,7, counterpart to \citet{Bhattacharya2017}'s X-ray source 371, presents clear variability within the 3 epochs of H$\alpha$ photometry, with 1 epoch presenting an emission in H$\alpha$, and the other 2  with no clear emission at all, showing that the mass transfer from source\,7 seems to vary in intensity with time. And if other sources present the same behaviour, we might be missing CVs only because they are in the lower mass transfer phase when they were observed in X-ray.  

Additionally, the lack of X-ray sources detected near most of our WD--MS targets does not completely rule out the possibility of some of those objects being CVs. A large portion of the non-magnetic CV population might be still undetected, because the X-ray luminosity of the majority of the CVs can be fainter than 4$\times10^{28}$\,erg\,s$^{-1}$ \citep{Pretorius2011}. At quiescence, the rate of material transferred onto the WD surface in dwarf novae (DNe) systems can be extremely low, making those objects spend most of their time at very faint X-ray luminosities \citep{Pretorius2011} and harder to detect. Furthermore, that could be one of the reasons for the scarce number of DNe detected in GCs \cite[e.g., one DNe and one DNe candidate in 47\,Tuc, ][]{Shara1996}. 

Another process that can interfere in the detection of X-ray luminosities in CVs is the magnetic field. Extremely strong magnetic fields observed in some CVs, such as AR UMa \cite[$\sim$230\,MG, ][]{Schmidt1996}, can reduce the accretion rate in these systems, making them spend most of their time ($\sim$80 per cent) in extended low states. This causes the X-ray luminosity in the low state to be 5 times fainter than the one presented in the high state \citep{Szkody1999}, making them difficult to detect.

One more possibility within the CV scope would be that some of our WD--MS systems could be so called detached CVs \citep{Davis2008,Zorotovic2016}. Such systems, once normal CVs, ceased to be so when their orbital period increased to about 3\,hours, stopping the mass transfer between the two stars. Those systems will appear as CVs again when the gravitational radiation that makes them evolve toward shorter periods causes them to reach an orbital period of $\approx$2\,h. At this point, the Roche lobe shrinks enough to restart mass transfer and the systems appear again as CVs at the lower edge of the period gap \citep{Rappaport1983}. However, we would need to determine the orbital period of our outliers to confirm if any of them fits in this category. 

The last point to discuss are our ELM candidates. Among the sources with no previous classification, we detected 2 ELM candidates, both presenting photometric behaviour consistent with ELMs models and have possible cluster membership. But whether they are part of a binary system or are, currently, single stars, is something that still has to be determined. Additionally, the nature of the possible companions to ELM candidates in GCs, is still undetermined. For the Galactic field, \cite{Brown2016} showed that the binary companions for most ELMs are canonical mass WDs, following a normal distribution with mean mass of 0.76$_{\odot}$ and dispersion of 0.25M$_{\odot}$. Likewise, in GCs, \cite{Hansen2003} inferred that the binary companions to ELMs in GCs are likely to be carbon-oxygen core WDs. However, a large portion of the ELM candidates detected in GCs were identified as companions to millisecond pulsars \citep[e.g.][]{Ferraro2003, Bassa2006}. Yet, the cluster membership of these objects has been neither confirmed nor ruled out \citep[e.g.][]{Bassa2006,Corongiu2006}.  So, whether the ELMs in GCs have neutron stars or massive WDs as companions is an open question that can only be answered with a systematic study of those objects in GCs. 

\section{Conclusions}
\label{conclusions}
We used multicolour photometry along with high-precision internal proper motions to identify a population of outliers from the globular cluster 47\,Tuc. The vast majority of the sources we reported here had never been catalogued before, opening the doors to the study of evolved binary systems in clusters with HST data.  

Among the 24 outliers with no previous classification, we were able to find 22 with no X-ray source detected nearby. Two outliers (8 and 12) are bluer than the main-sequence both in the UV and optical colours (i.e., non-shifters). While 22 outliers with no previous classification are shifters, they are bluer than the main-sequence in the UV CMD, but shift to the main-sequence, or redder colour, in the visual CMD. 

In our proper-motion selected sample, we found 7 outliers with X-ray sources nearby. Five have some previous classification, most of them as CVs or CV candidates. The remaining 2 outliers, both shifters, present an X-ray source detected within 0.4\,arcsecond. All these clues point to the idea that those sources are likely some kind of interacting binaries, going through mass transfer. Along with that, One of the outliers with X-ray counterpart, source\,7, presents clear variability within the 3 epochs of H$\alpha$ photometry, with 1 epoch presenting an emission in H$\alpha$, and the other 2  with no clear emission at all, indicating that the mass transfer from source\,7 varies in intensity with time.  

When comparing the positions of our non-shifters in the visual and UV CMDs to the evolutionary models from \cite{Istrate2016}, we identify that sources 8 and 12 can be ELMs in the proto-WD phase. However, they  could also be possible ELMs with masses lower than 0.18\,M$_{\odot}$ with a formation channel possibly different than the stable Roche-lobe overflow mass transfer. We also explored the possibility of outlier 26, previously classified as a possible W UMa binary, being a double-degenerate system, composed of an ELM and a massive WD. Outlier\,26 is very close to the $\sim0.18\,M_{\odot}$ model in both visual and UV CMDs, and its variability period is consistent with the ones found for ELMs in the Galactic field with the same mass. Furthermore, sources 8, 12 and 26  are all found to have significant H$\alpha$-absorption, as predicted for ELMs. 

The current knowledge we have on the outliers allows us only to speculate about their true nature. Multicolour photometry has guided their discovery. However, only with detailed studies of those objects we can improve our understanding of binary interactions in regimes of parameters and dynamics different from the ones in the Galactic field. However, that can only be achieved with additional data to clarify in what specific category each of our outliers fit. Light curves, along with deeper spectroscopic and X-ray data, are necessary in order to precisely classify these objects and enhance our knowledge of binary stellar-evolution.

\section*{Acknowledgements}

We would like to thank the anonymous referee for the constructive comments, which helped us to improve the manuscript. F.C would like to thank Dr. Adrienne M. Cool for constructive criticism of the manuscript. Based on observations with the NASA/ESA Hubble Space Telescope, obtained at the Space Telescope Science Institute, which is operated by AURA, Inc., under NASA contract NAS 5-26555. F.C. was partially supported by NSF grant AST-1211616. I.P. was supported by DFG under grant GE 2056-12-1. S.O.K. was supported by CNPq-Brazil. S.K. gratefully acknowledges support from the European Research Council (ERC-CoG-646928, Multi-Pop). S.D., S.K., and T.O.H. acknowledge support from the German Ministry for Education and Science (BMBF Verbundforschung) through grants 05A14MGA, 05A17MGA, 05A14BAC, and 05A17BAA.This research is supported by the German Research Foundation (DFG) with grants DR 281/35-1 and KA 4537/2-1. A.B. acknowledges support from HST grant AR-12845 and GO-13297. D.N. and G.P. acknowledge partial support by the Universit\`a degli Studi di Padova Progetto di Ateneo CPDA141214 and BIRD178590 and by INAF under the program PRIN-INAF2014. M.H.M. and D.E.W. acknowledge support from the United States Department of Energy under grant DE-SC0010623, the Wooton Center for Astrophysical Plasma Properties under the United States Department of Energy under grant DE-FOA-0001634, and the NSF grant AST 1707419. A.D. was supported by NSF grant AST-1313280. A.G.I. acknowledges support from the NASA Astrophysics Theory Program through NASA grant NNX13AH43G. 



\bibliographystyle{mnras}
\bibliography{references} 

\begin{thebibliography}{}
\makeatletter
\relax
\def\mn@urlcharsother{\let\do\@makeother \do\$\do\&\do\#\do\^\do\_\do\%\do\~}
\def\mn@doi{\begingroup\mn@urlcharsother \@ifnextchar [ {\mn@doi@}
  {\mn@doi@[]}}
\def\mn@doi@[#1]#2{\def\@tempa{#1}\ifx\@tempa\@empty \href
  {http://dx.doi.org/#2} {doi:#2}\else \href {http://dx.doi.org/#2} {#1}\fi
  \endgroup}
\def\mn@eprint#1#2{\mn@eprint@#1:#2::\@nil}
\def\mn@eprint@arXiv#1{\href {http://arxiv.org/abs/#1} {{\tt arXiv:#1}}}
\def\mn@eprint@dblp#1{\href {http://dblp.uni-trier.de/rec/bibtex/#1.xml}
  {dblp:#1}}
\def\mn@eprint@#1:#2:#3:#4\@nil{\def\@tempa {#1}\def\@tempb {#2}\def\@tempc
  {#3}\ifx \@tempc \@empty \let \@tempc \@tempb \let \@tempb \@tempa \fi \ifx
  \@tempb \@empty \def\@tempb {arXiv}\fi \@ifundefined
  {mn@eprint@\@tempb}{\@tempb:\@tempc}{\expandafter \expandafter \csname
  mn@eprint@\@tempb\endcsname \expandafter{\@tempc}}}

\bibitem[\protect\citeauthoryear{Albrow, Gilliland, Brown, Edmonds,
  Guhathakurta  \& Sarajedini}{Albrow et~al.}{2001}]{Albrow2001}
Albrow M.~D.,  Gilliland R.~L.,  Brown T.~M.,  Edmonds P.~D.,  Guhathakurta P.,
    Sarajedini A.,  2001, \mn@doi [\aj] {10.1086/322353}, 559, 1060

\bibitem[\protect\citeauthoryear{Anderson \& King}{Anderson \&
  King}{2006}]{Anderson2006}
Anderson J.,  King I.~R.,  2006, Technical report, PSFs, Photometry, and
  Astronomy for the ACS/WFC.
STSCI

\bibitem[\protect\citeauthoryear{Bacon et~al.,}{Bacon et~al.}{2010}]{Bacon2010}
Bacon R.,  et~al., 2010, in Ground-based and Airborne Instrumentation for
  Astronomy III. p. 773508

\bibitem[\protect\citeauthoryear{Bassa, van Kerkwijk, Koester  \&
  Verbunt}{Bassa et~al.}{2006}]{Bassa2006}
Bassa C.~G.,  van Kerkwijk M.~H.,  Koester D.,   Verbunt F.,  2006, \mn@doi
  [\aap] {10.1051/0004-6361:20065181}, 456, 295

\bibitem[\protect\citeauthoryear{Bellini \& Bedin}{Bellini \&
  Bedin}{2009}]{Bellini2009}
Bellini A.,  Bedin L.~R.,  2009, \mn@doi [\pasp] {10.1086/649061}, 121, 1419

\bibitem[\protect\citeauthoryear{Bellini, Anderson  \& Bedin}{Bellini
  et~al.}{2011}]{Bellini2011}
Bellini A.,  Anderson J.,   Bedin L.~R.,  2011, \mn@doi [\pasp]
  {10.1086/659878}, 123, 622

\bibitem[\protect\citeauthoryear{Bellini, Anderson, Salaris, Cassisi, Bedin,
  Piotto  \& Bergeron}{Bellini et~al.}{2013}]{Bellini2013}
Bellini A.,  Anderson J.,  Salaris M.,  Cassisi S.,  Bedin L.~R.,  Piotto G.,
  Bergeron P.,  2013, \mn@doi [\apj] {10.1088/2041-8205/769/2/l32}, 769, L32

\bibitem[\protect\citeauthoryear{Bellini et~al.,}{Bellini
  et~al.}{2014}]{Bellini2014}
Bellini A.,  et~al., 2014, \mn@doi [\apj] {10.1088/0004-637X/797/2/115}, 797,
  115

\bibitem[\protect\citeauthoryear{Belloni, Giersz, Askar, Leigh  \&
  Hypki}{Belloni et~al.}{2016}]{Belloni2016}
Belloni D.,  Giersz M.,  Askar A.,  Leigh N.,   Hypki A.,  2016, \mn@doi
  [\mnras] {10.1093/mnras/stw1841}, 462, 2950

\bibitem[\protect\citeauthoryear{Benacquista \& Downing}{Benacquista \&
  Downing}{2013}]{Benacquista2013}
Benacquista M.~J.,  Downing J. M.~B.,  2013, \mn@doi [Living Reviews in
  Relativity] {10.12942/lrr-2013-4}, 16

\bibitem[\protect\citeauthoryear{Bhattacharya, Heinke, Chugunov, Freire,
  Ridolfi  \& Bogdanov}{Bhattacharya et~al.}{2017}]{Bhattacharya2017}
Bhattacharya S.,  Heinke C.~O.,  Chugunov A.~I.,  Freire P. C.~C.,  Ridolfi A.,
    Bogdanov S.,  2017, \mn@doi [\mnras] {10.1093/mnras/stx2241}, 472, 3706

\bibitem[\protect\citeauthoryear{Brown, Kilic, Allende~Prieto  \& Kenyon}{Brown
  et~al.}{2010}]{Brown2010}
Brown W.~R.,  Kilic M.,  Allende~Prieto C.,   Kenyon S.~J.,  2010, \mn@doi
  [\apj] {10.1088/0004-637X/723/2/1072}, 723, 1072

\bibitem[\protect\citeauthoryear{Brown, Gianninas, Kilic, Kenyon  \&
  Prieto}{Brown et~al.}{2016}]{Brown2016}
Brown W.~R.,  Gianninas A.,  Kilic M.,  Kenyon S.~J.,   Prieto C.~A.,  2016,
  \mn@doi [\apj] {10.3847/0004-637x/818/2/155}, 818, 155

\bibitem[\protect\citeauthoryear{Campos, Kepler, Bonatto  \& Ducati}{Campos
  et~al.}{2013}]{Campos2013}
Campos F.,  Kepler S.~O.,  Bonatto C.,   Ducati J.~R.,  2013, \mn@doi [\mnras]
  {10.1093/mnras/stt719}, 433, 243

\bibitem[\protect\citeauthoryear{Campos et~al.,}{Campos
  et~al.}{2016}]{Campos2016}
Campos F.,  et~al., 2016, \mn@doi [\mnras] {10.1093/mnras/stv2911}, 456, 3729

\bibitem[\protect\citeauthoryear{Chen, Girardi, Bressan, Marigo, Barbieri  \&
  Kong}{Chen et~al.}{2014}]{Chen2014}
Chen Y.,  Girardi L.,  Bressan A.,  Marigo P.,  Barbieri M.,   Kong X.,  2014,
  \mn@doi [\mnras] {10.1093/mnras/stu1605}, 444, 2525

\bibitem[\protect\citeauthoryear{Choi, Dotter, Conroy, Cantiello, Paxton  \&
  Johnson}{Choi et~al.}{2016}]{Choi2016}
Choi J.,  Dotter A.,  Conroy C.,  Cantiello M.,  Paxton B.,   Johnson B.~D.,
  2016, \mn@doi [\apj] {10.3847/0004-637x/823/2/102}, 823, 102

\bibitem[\protect\citeauthoryear{Clement et~al.,}{Clement
  et~al.}{2001}]{Clement2001}
Clement C.~M.,  et~al., 2001, \mn@doi [\apj] {10.1086/323719}, 122, 2587

\bibitem[\protect\citeauthoryear{Cool, Grindlay, Cohn, Lugger  \& Bailyn}{Cool
  et~al.}{1998}]{Cool1998}
Cool A.~M.,  Grindlay J.~E.,  Cohn H.~N.,  Lugger P.~M.,   Bailyn C.~D.,  1998,
  \mn@doi [\apjl] {10.1086/311730}, 508, L75

\bibitem[\protect\citeauthoryear{Corongiu, Possenti, Lyne, Manchester, Camilo,
  D{\textquotesingle}Amico  \& Sarkissian}{Corongiu
  et~al.}{2006}]{Corongiu2006}
Corongiu A.,  Possenti A.,  Lyne A.~G.,  Manchester R.~N.,  Camilo F.,
  D{\textquotesingle}Amico N.,   Sarkissian J.~M.,  2006, \mn@doi [\apj]
  {10.1086/508918}, 653, 1417

\bibitem[\protect\citeauthoryear{Davies}{Davies}{2002}]{Davies2002}
Davies M.~B.,  2002, in {van Leeuwen} F.,  {Hughes} J.~D.,   {Piotto} G.,  eds,
  ~ASPC Vol. 265, \aspc. p.~215

\bibitem[\protect\citeauthoryear{Davis, Kolb, Willems  \& G{\"a}nsicke}{Davis
  et~al.}{2008}]{Davis2008}
Davis P.~J.,  Kolb U.,  Willems B.,   G{\"a}nsicke B.~T.,  2008, \mn@doi
  [\mnras] {10.1111/j.1365-2966.2008.13675.x}, 389, 1563

\bibitem[\protect\citeauthoryear{Dotter}{Dotter}{2016}]{Dotter2016}
Dotter A.,  2016, \mn@doi [\aaps] {10.3847/0067-0049/222/1/8}, 222, 8

\bibitem[\protect\citeauthoryear{Edmonds, Gilliland, Heinke  \&
  Grindlay}{Edmonds et~al.}{2003}]{Edmonds2003}
Edmonds P.~D.,  Gilliland R.~L.,  Heinke C.~O.,   Grindlay J.~E.,  2003,
  \mn@doi [\apj] {10.1086/378193}, 596, 1177

\bibitem[\protect\citeauthoryear{Eyer et~al.,}{Eyer et~al.}{2018}]{Eyer2018}
Eyer L.,  et~al., 2018, \aap, arXiv:1804.09382

\bibitem[\protect\citeauthoryear{Fabian, Pringle  \& Rees}{Fabian
  et~al.}{1975}]{Fabian1975}
Fabian A.~C.,  Pringle J.~E.,   Rees M.~J.,  1975, \mn@doi [\mnras]
  {10.1093/mnras/172.1.15p}, 172, 15P

\bibitem[\protect\citeauthoryear{Ferraro, Possenti, Sabbi  \&
  D{\textquotesingle}Amico}{Ferraro et~al.}{2003}]{Ferraro2003}
Ferraro F.~R.,  Possenti A.,  Sabbi E.,   D{\textquotesingle}Amico N.,  2003,
  \mn@doi [\apj] {10.1086/379536}, 596, L211

\bibitem[\protect\citeauthoryear{Fontaine, Brassard  \& Bergeron}{Fontaine
  et~al.}{2001}]{Fontaine2001}
Fontaine G.,  Brassard P.,   Bergeron P.,  2001, \mn@doi [\pasp]
  {10.1086/319535}, 113, 409

\bibitem[\protect\citeauthoryear{Freire et~al.,}{Freire
  et~al.}{2017}]{Freire2017}
Freire P. C.~C.,  et~al., 2017, \mn@doi [\mnras] {10.1093/mnras/stx1533}, 471,
  857

\bibitem[\protect\citeauthoryear{Garcia, McClintock, Narayan, Callanan, Barret
  \& Murray}{Garcia et~al.}{2001}]{Garcia2001}
Garcia M.~R.,  McClintock J.~E.,  Narayan R.,  Callanan P.,  Barret D.,
  Murray S.~S.,  2001, \mn@doi [\apj] {10.1086/320494}, 553, L47

\bibitem[\protect\citeauthoryear{Giersz \& Spurzem}{Giersz \&
  Spurzem}{2000}]{Giersz2000}
Giersz M.,  Spurzem R.,  2000, \mn@doi [\mnras]
  {10.1046/j.1365-8711.2000.03727.x}, 317, 581

\bibitem[\protect\citeauthoryear{Giesers et~al.,}{Giesers
  et~al.}{2018}]{Giesers2018}
Giesers B.,  et~al., 2018, \mn@doi [\mnras] {10.1093/mnrasl/slx203}, 475, L15

\bibitem[\protect\citeauthoryear{Goodman \& Hut}{Goodman \&
  Hut}{1989}]{Goodman1989}
Goodman J.,  Hut P.,  1989, \mn@doi [\nat] {10.1038/339040a0}, 339, 40

\bibitem[\protect\citeauthoryear{Hansen, Kalogera  \& Rasio}{Hansen
  et~al.}{2003}]{Hansen2003}
Hansen B. M.~S.,  Kalogera V.,   Rasio F.~A.,  2003, \mn@doi [\apj]
  {10.1086/367827}, 586, 1364

\bibitem[\protect\citeauthoryear{Harris}{Harris}{1996}]{Harris1996}
Harris W.~E.,  1996, \mn@doi [\aj] {10.1086/118116}, 112, 1487

\bibitem[\protect\citeauthoryear{Haurberg, Lubell, Cohn, Lugger, Anderson, Cool
   \& Serenelli}{Haurberg et~al.}{2010}]{Haurberg2010}
Haurberg N.~C.,  Lubell G. M.~G.,  Cohn H.~N.,  Lugger P.~M.,  Anderson J.,
  Cool A.~M.,   Serenelli A.~M.,  2010, \mn@doi [\apj]
  {10.1088/0004-637x/722/1/158}, 722, 158

\bibitem[\protect\citeauthoryear{Heggie}{Heggie}{1975}]{Heggie1975}
Heggie D.~C.,  1975, \mn@doi [\mnras] {10.1093/mnras/173.3.729}, 173, 729

\bibitem[\protect\citeauthoryear{Heggie \& Aarseth}{Heggie \&
  Aarseth}{1992}]{Heggie1992}
Heggie D.~C.,  Aarseth S.~J.,  1992, \mn@doi [\mnras]
  {10.1093/mnras/257.3.513}, 257, 513

\bibitem[\protect\citeauthoryear{Heggie, Trenti  \& Hut}{Heggie
  et~al.}{2006}]{Heggie2006}
Heggie D.~C.,  Trenti M.,   Hut P.,  2006, \mn@doi [\mnras]
  {10.1111/j.1365-2966.2006.10122.x}, 368, 677

\bibitem[\protect\citeauthoryear{Heinke, Grindlay, Edmonds, Cohn, Lugger,
  Camilo, Bogdanov  \& Freire}{Heinke et~al.}{2005}]{Heinke2005}
Heinke C.~O.,  Grindlay J.~E.,  Edmonds P.~D.,  Cohn H.~N.,  Lugger P.~M.,
  Camilo F.,  Bogdanov S.,   Freire P.~C.,  2005, \mn@doi [\aj]
  {10.1086/429899}, 625, 796

\bibitem[\protect\citeauthoryear{Helmi et~al.,}{Helmi et~al.}{2018}]{Helmi2018}
Helmi A.,  et~al., 2018, \mn@doi [\aap] {10.1051/0004-6361/201832698}, 616, A12

\bibitem[\protect\citeauthoryear{Henleywillis, Cool, Haggard, Heinke, Callanan
  \& Zhao}{Henleywillis et~al.}{2018}]{Henleywillis2018}
Henleywillis S.,  Cool A.~M.,  Haggard D.,  Heinke C.,  Callanan P.,   Zhao Y.,
   2018, \mn@doi [\mnras] {10.1093/mnras/sty675}

\bibitem[\protect\citeauthoryear{Heyl, Caiazzo, Richer, Anderson, Kalirai  \&
  Parada}{Heyl et~al.}{2017}]{Heyl2017}
Heyl J.,  Caiazzo I.,  Richer H.,  Anderson J.,  Kalirai J.,   Parada J.,
  2017, \mn@doi [\apj] {10.3847/1538-4357/aa974f}, 850, 186

\bibitem[\protect\citeauthoryear{Hong, Vesperini, Belloni  \& Giersz}{Hong
  et~al.}{2017}]{Hong2017}
Hong J.,  Vesperini E.,  Belloni D.,   Giersz M.,  2017, \mn@doi [\mnras]
  {10.1093/mnras/stw2595}, 464, 2511

\bibitem[\protect\citeauthoryear{Hurley \& Tout}{Hurley \&
  Tout}{1998}]{Hurley1998}
Hurley J.,  Tout C.~A.,  1998, \mn@doi [\mnras]
  {10.1046/j.1365-8711.1998.01981.x}, 300, 977

\bibitem[\protect\citeauthoryear{Hut \& Bahcall}{Hut \&
  Bahcall}{1983}]{Hut1983}
Hut P.,  Bahcall J.~N.,  1983, \mn@doi [\apj] {10.1086/160956}, 268, 319

\bibitem[\protect\citeauthoryear{Hut \& Verbunt}{Hut \&
  Verbunt}{1983}]{Hut1983a}
Hut P.,  Verbunt F.,  1983, \mn@doi [\nat] {10.1038/301587a0}, 301, 587

\bibitem[\protect\citeauthoryear{Hut, McMillan  \& Romani}{Hut
  et~al.}{1992}]{Hut1992}
Hut P.,  McMillan S.,   Romani R.~W.,  1992, \mn@doi [\apj] {10.1086/171229},
  389, 527

\bibitem[\protect\citeauthoryear{Istrate, Tauris  \& Langer}{Istrate
  et~al.}{2014}]{Istrate2014}
Istrate A.~G.,  Tauris T.~M.,   Langer N.,  2014, \mn@doi [\aap]
  {10.1051/0004-6361/201424680}, 571, A45

\bibitem[\protect\citeauthoryear{Istrate, Marchant, Tauris, Langer, Stancliffe
  \& Grassitelli}{Istrate et~al.}{2016}]{Istrate2016}
Istrate A.~G.,  Marchant P.,  Tauris T.~M.,  Langer N.,  Stancliffe R.~J.,
  Grassitelli L.,  2016, \mn@doi [\aap] {10.1051/0004-6361/201628874}, 595, A35

\bibitem[\protect\citeauthoryear{Ivanova, Belczynski, Fregeau  \&
  Rasio}{Ivanova et~al.}{2005}]{Ivanova2005}
Ivanova N.,  Belczynski K.,  Fregeau J.~M.,   Rasio F.~A.,  2005, \mn@doi
  [\mnras] {10.1111/j.1365-2966.2005.08804.x}, 358, 572

\bibitem[\protect\citeauthoryear{Ivanova, Heinke, Rasio, Taam, Belczynski  \&
  Fregeau}{Ivanova et~al.}{2006}]{Ivanova2006}
Ivanova N.,  Heinke C.~O.,  Rasio F.~A.,  Taam R.~E.,  Belczynski K.,   Fregeau
  J.,  2006, \mn@doi [\mnras] {10.1111/j.1365-2966.2006.10876.x}, 372, 1043

\bibitem[\protect\citeauthoryear{Kalirai, Davis, Richer, Bergeron, Catelan,
  Hansen  \& Rich}{Kalirai et~al.}{2009}]{Kalirai2009}
Kalirai J.~S.,  Davis D.~S.,  Richer H.~B.,  Bergeron P.,  Catelan M.,  Hansen
  B. M.~S.,   Rich R.~M.,  2009, \mn@doi [\apj] {10.1088/0004-637x/705/1/408},
  705, 408

\bibitem[\protect\citeauthoryear{Kamann et~al.,}{Kamann
  et~al.}{2018}]{Kamann2017}
Kamann S.,  et~al., 2018, \mn@doi [\mnras] {10.1093/mnras/stx2719}, 473, 5591

\bibitem[\protect\citeauthoryear{Knigge}{Knigge}{2011}]{Knigge2011}
Knigge C.,  2011, in Evolution of Compact Binaries. p.~3

\bibitem[\protect\citeauthoryear{Knigge, Dieball, Ma{\'{\i}}z~Apell{\'a}niz,
  Long, Zurek  \& Shara}{Knigge et~al.}{2008}]{Knigge2008}
Knigge C.,  Dieball A.,  Ma{\'{\i}}z~Apell{\'a}niz J.,  Long K.~S.,  Zurek
  D.~R.,   Shara M.~M.,  2008, \mn@doi [\apj] {10.1086/589987}, 683, 1006

\bibitem[\protect\citeauthoryear{Kouwenhoven, Brown, Zinnecker, Kaper  \&
  Zwart}{Kouwenhoven et~al.}{2005}]{Kouwenhoven2005}
Kouwenhoven M. B.~N.,  Brown A. G.~A.,  Zinnecker H.,  Kaper L.,   Zwart S.
  F.~P.,  2005, \mn@doi [\aap] {10.1051/0004-6361:20048124}, 430, 137

\bibitem[\protect\citeauthoryear{Leigh, Giersz, Marks, Webb, Hypki, Heinke,
  Kroupa  \& Sills}{Leigh et~al.}{2015}]{Leigh2015}
Leigh N. W.~C.,  Giersz M.,  Marks M.,  Webb J.~J.,  Hypki A.,  Heinke C.~O.,
  Kroupa P.,   Sills A.,  2015, \mn@doi [\mnras] {10.1093/mnras/stu2110}, 446,
  226

\bibitem[\protect\citeauthoryear{Liu, van Paradijs  \& van~den Heuvel}{Liu
  et~al.}{2007}]{Liu2007}
Liu Q.~Z.,  van Paradijs J.,   van~den Heuvel E. P.~J.,  2007, \mn@doi [\aap]
  {10.1051/0004-6361:20077303}, 469, 807

\bibitem[\protect\citeauthoryear{Lorimer}{Lorimer}{2008}]{Lorimer2008}
Lorimer D.~R.,  2008, \mn@doi [LRR] {10.12942/lrr-2008-8}, 11

\bibitem[\protect\citeauthoryear{McConnell \& Ables}{McConnell \&
  Ables}{2000}]{McConnell2000}
McConnell D.,  Ables J.~G.,  2000, \mn@doi [\mnras]
  {10.1046/j.1365-8711.2000.03157.x}, 311, 841

\bibitem[\protect\citeauthoryear{McMillan}{McMillan}{1986}]{McMillan1986}
McMillan S. L.~W.,  1986, \mn@doi [\apj] {10.1086/164365}, 306, 552

\bibitem[\protect\citeauthoryear{McMillan, McDermott  \& Taam}{McMillan
  et~al.}{1987}]{McMillan1987}
McMillan S. L.~W.,  McDermott P.~N.,   Taam R.~E.,  1987, \mn@doi [\apj]
  {10.1086/165365}, 318, 261

\bibitem[\protect\citeauthoryear{Miller-Jones et~al.,}{Miller-Jones
  et~al.}{2015}]{Miller-Jones2015}
Miller-Jones J. C.~A.,  et~al., 2015, \mn@doi [\mnras] {10.1093/mnras/stv1869},
  453, 3919

\bibitem[\protect\citeauthoryear{Milone et~al.,}{Milone
  et~al.}{2012}]{Milone2012}
Milone A.~P.,  et~al., 2012, \mn@doi [\aap] {10.1051/0004-6361/201016384}, 540,
  A16

\bibitem[\protect\citeauthoryear{Moehler, Koester, Zoccali, Ferraro, Heber,
  Napiwotzki  \& Renzini}{Moehler et~al.}{2004}]{Moehler2004}
Moehler S.,  Koester D.,  Zoccali M.,  Ferraro F.~R.,  Heber U.,  Napiwotzki
  R.,   Renzini A.,  2004, \mn@doi [\aap] {10.1051/0004-6361:20035819}, 420,
  515

\bibitem[\protect\citeauthoryear{Narayan \& McClintock}{Narayan \&
  McClintock}{2008}]{Narayan2008}
Narayan R.,  McClintock J.~E.,  2008, \mn@doi [New Astronomy Reviews]
  {10.1016/j.newar.2008.03.002}, 51, 733

\bibitem[\protect\citeauthoryear{Narayan, Barret  \& McClintock}{Narayan
  et~al.}{1997}]{Narayan1997}
Narayan R.,  Barret D.,   McClintock J.~E.,  1997, \mn@doi [\apj]
  {10.1086/304134}, 482, 448

\bibitem[\protect\citeauthoryear{Nardiello et~al.,}{Nardiello
  et~al.}{2018}]{Nardiello2018}
Nardiello D.,  et~al., 2018, \mnras, arXiv:1809.04300

\bibitem[\protect\citeauthoryear{Nelemans \& Jonker}{Nelemans \&
  Jonker}{2010}]{Nelemans2010}
Nelemans G.,  Jonker P.,  2010, \mn@doi [New Astronomy Reviews]
  {10.1016/j.newar.2010.09.021}, 54, 87

\bibitem[\protect\citeauthoryear{Piotto et~al.,}{Piotto
  et~al.}{2015}]{Piotto2015}
Piotto G.,  et~al., 2015, \mn@doi [\aj] {10.1088/0004-6256/149/3/91}, 149, 91

\bibitem[\protect\citeauthoryear{Pooley}{Pooley}{2010}]{Pooley2010}
Pooley D.,  2010, \mn@doi [PNAS] {10.1073/pnas.0913903107}, 107, 7164

\bibitem[\protect\citeauthoryear{Pretorius \& Knigge}{Pretorius \&
  Knigge}{2011}]{Pretorius2011}
Pretorius M.~L.,  Knigge C.,  2011, \mn@doi [\mnras]
  {10.1111/j.1365-2966.2011.19801.x}, 419, 1442

\bibitem[\protect\citeauthoryear{Rappaport, Joss  \& Verbunt}{Rappaport
  et~al.}{1983}]{Rappaport1983}
Rappaport S.,  Joss P.~C.,   Verbunt F.,  1983, \mn@doi [\apj]
  {10.1086/161569}, 275, 713

\bibitem[\protect\citeauthoryear{Ridolfi et~al.,}{Ridolfi
  et~al.}{2016}]{Ridolfi2016}
Ridolfi A.,  et~al., 2016, \mn@doi [\mnras] {10.1093/mnras/stw1850}, 462, 2918

\bibitem[\protect\citeauthoryear{{Rivera Sandoval} et~al.,}{{Rivera Sandoval}
  et~al.}{2018}]{Sandoval2018}
{Rivera Sandoval} L.~E.,  et~al., 2018, \mn@doi [\mnras]
  {10.1093/mnras/sty058}, 475, 4841

\bibitem[\protect\citeauthoryear{Robin, Reyl{\'{e}}, Derri{\`{e}}re  \&
  Picaud}{Robin et~al.}{2003}]{Robin2003}
Robin A.~C.,  Reyl{\'{e}} C.,  Derri{\`{e}}re S.,   Picaud S.,  2003, \mn@doi
  [\aap] {10.1051/0004-6361:20031117}, 409, 523

\bibitem[\protect\citeauthoryear{Rucinski}{Rucinski}{1994}]{Rucinski1994}
Rucinski S.~M.,  1994, \mn@doi [\pasp] {10.1086/133401}, 106, 462

\bibitem[\protect\citeauthoryear{Rucinski}{Rucinski}{1995}]{Rucinski1995}
Rucinski S.,  1995, \mn@doi [\pasp] {10.1086/133603}, 107, 648

\bibitem[\protect\citeauthoryear{Salaris, Cassisi, Pietrinferni, Kowalski  \&
  Isern}{Salaris et~al.}{2010}]{Salaris2010}
Salaris M.,  Cassisi S.,  Pietrinferni A.,  Kowalski P.~M.,   Isern J.,  2010,
  \mn@doi [\apj] {10.1088/0004-637x/716/2/1241}, 716, 1241

\bibitem[\protect\citeauthoryear{Schmidt, Szkody, Smith, Silber, Tovmassian,
  Hoard, Gansicke  \& de Martino}{Schmidt et~al.}{1996}]{Schmidt1996}
Schmidt G.~D.,  Szkody P.,  Smith P.~S.,  Silber A.,  Tovmassian G.,  Hoard
  D.~W.,  Gansicke B.~T.,   de Martino D.,  1996, \mn@doi [\apj]
  {10.1086/178160}, 473, 483

\bibitem[\protect\citeauthoryear{Shara, Bergeron, Gilliland, Saha  \&
  Petro}{Shara et~al.}{1996}]{Shara1996}
Shara M.~M.,  Bergeron L.~E.,  Gilliland R.~L.,  Saha A.,   Petro L.,  1996,
  \mn@doi [\apj] {10.1086/178008}, 471, 804

\bibitem[\protect\citeauthoryear{Shull}{Shull}{1979}]{Shull1979}
Shull J.~M.,  1979, \mn@doi [\apj] {10.1086/157215}, 231, 534

\bibitem[\protect\citeauthoryear{Sollima, Beccari, Ferraro, Pecci  \&
  Sarajedini}{Sollima et~al.}{2007}]{Sollima2007}
Sollima A.,  Beccari G.,  Ferraro F.~R.,  Pecci F.~F.,   Sarajedini A.,  2007,
  \mn@doi [\mnras] {10.1111/j.1365-2966.2007.12116.x}, 380, 781

\bibitem[\protect\citeauthoryear{Strickler, Cool, Anderson, Cohn, Lugger  \&
  Serenelli}{Strickler et~al.}{2009}]{Strickler2009}
Strickler R.~R.,  Cool A.~M.,  Anderson J.,  Cohn H.~N.,  Lugger P.~M.,
  Serenelli A.~M.,  2009, \mn@doi [\apj] {10.1088/0004-637x/699/1/40}, 699, 40

\bibitem[\protect\citeauthoryear{Szkody, Vennes, Schmidt, Wagner, Fried,
  Shafter  \& Fierce}{Szkody et~al.}{1999}]{Szkody1999}
Szkody P.,  Vennes S.,  Schmidt G.~D.,  Wagner R.~M.,  Fried R.,  Shafter
  A.~W.,   Fierce E.,  1999, \mn@doi [\apj] {10.1086/307505}, 520, 841

\bibitem[\protect\citeauthoryear{Tanikawa, Hut  \& Makino}{Tanikawa
  et~al.}{2012}]{Tanikawa2012}
Tanikawa A.,  Hut P.,   Makino J.,  2012, \mn@doi [New Astronomy]
  {10.1016/j.newast.2011.09.001}, 17, 272

\bibitem[\protect\citeauthoryear{Taylor, Grindlay, Edmonds  \& Cool}{Taylor
  et~al.}{2001}]{Taylor2001}
Taylor J.~M.,  Grindlay J.~E.,  Edmonds P.~D.,   Cool A.~M.,  2001, \mn@doi
  [\apj] {10.1086/320676}, 553, L169

\bibitem[\protect\citeauthoryear{Zorotovic et~al.,}{Zorotovic
  et~al.}{2016}]{Zorotovic2016}
Zorotovic M.,  et~al., 2016, \mn@doi [\mnras] {10.1093/mnras/stw246}, 457, 3867

\makeatother
\end{thebibliography}





\appendix
\section{Finding charts}
\label{Appendix A}
We present finding charts in F814W (ACS) and F336W (WFC3) for our 24 outliers with no previous classification listed in table tab. \ref{allout}. We also include the finding chart for source 26. The images are about 4$\times$5\,arcsec.

\begin{figure*}
\center
 \includegraphics[width=\textwidth]{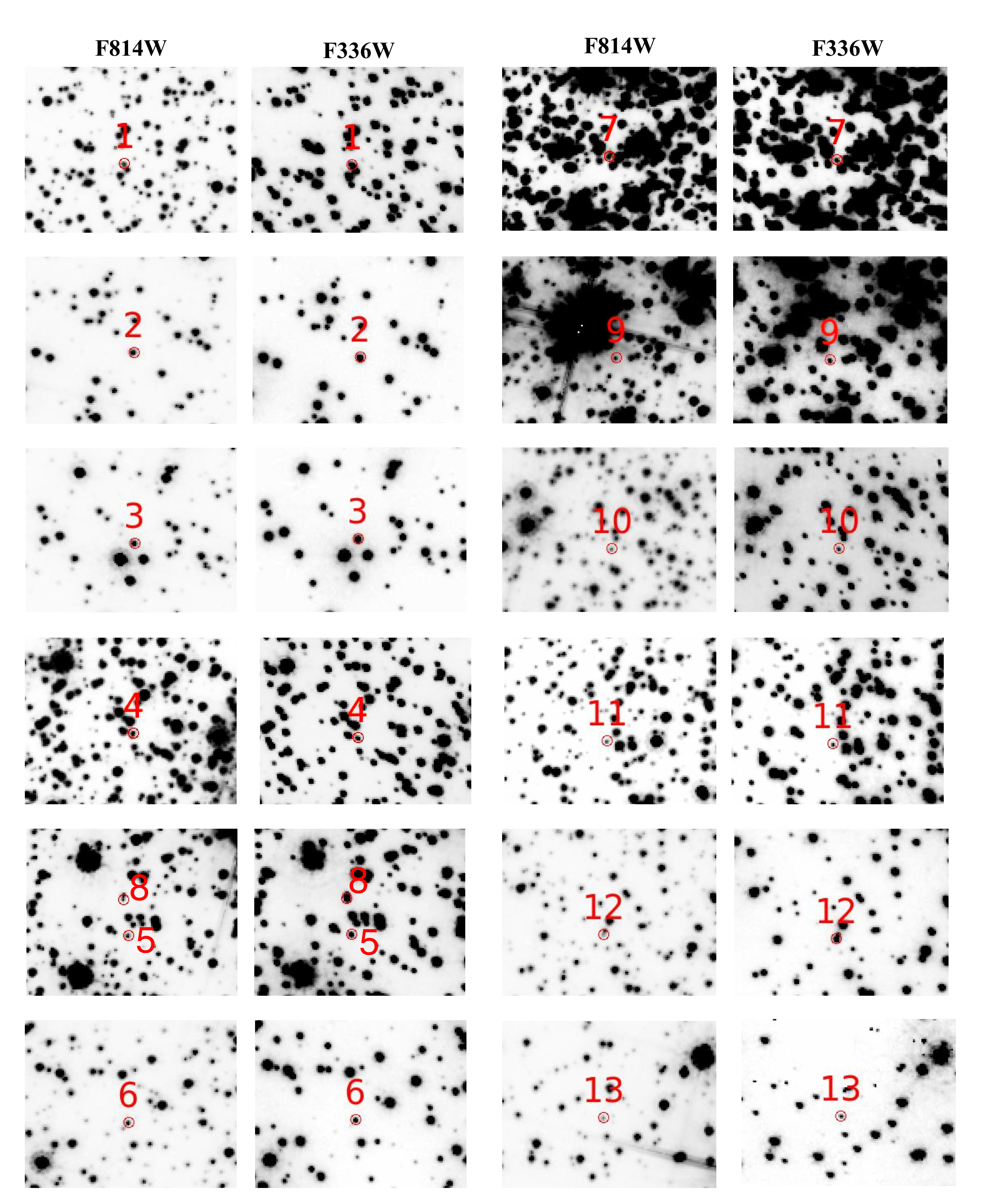}
 \caption{Finding charts in the filters F814W and F336W for outliers 1--13
}
 \label{find1}
\end{figure*}

\begin{figure*}
\center
 \includegraphics[width=\textwidth]{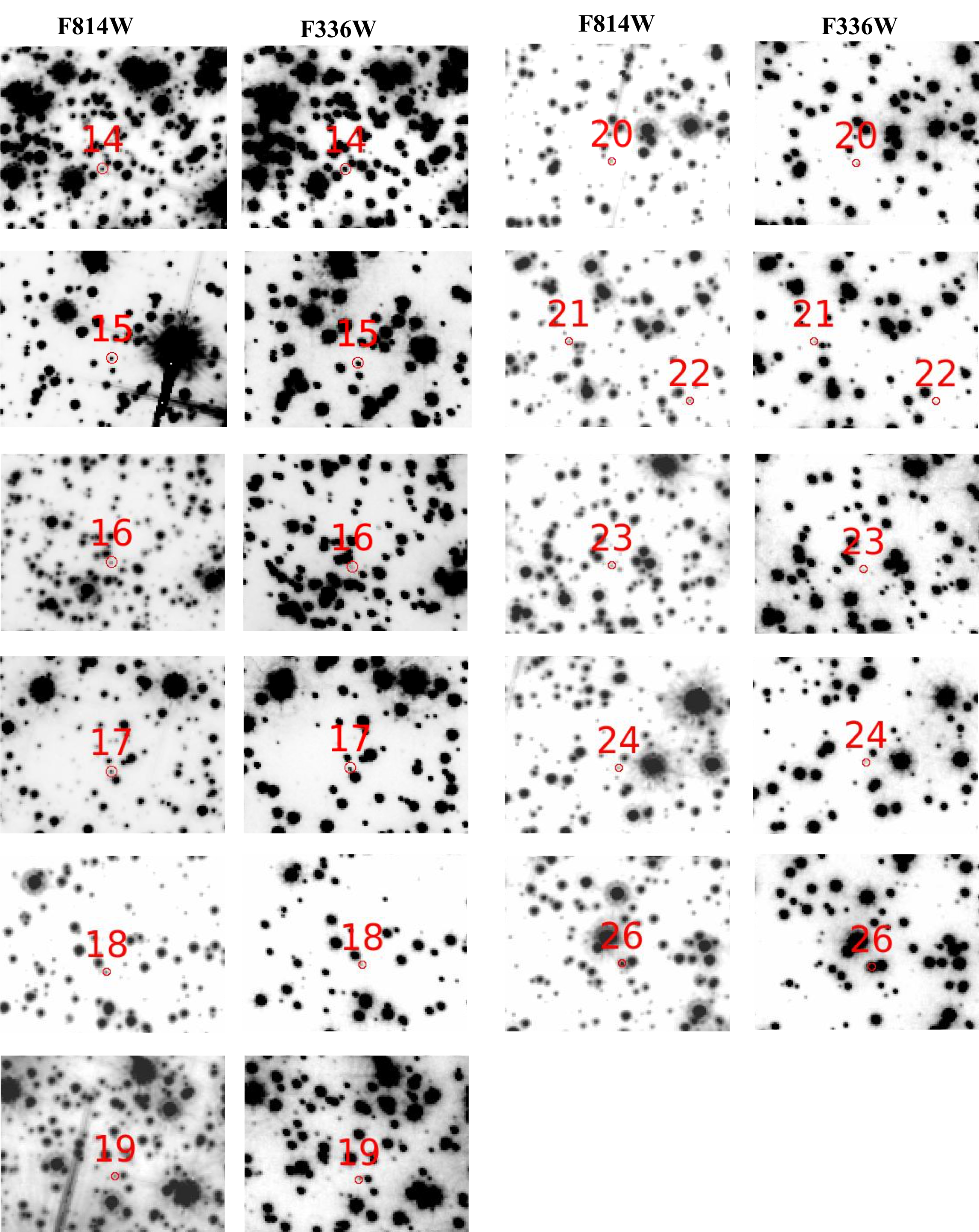}
 \caption{Finding charts in the filters F814W and F336W for outliers 14--24 and source 26.
}
 \label{find2}
\end{figure*}


\bsp	
\label{lastpage}
\end{document}